\documentclass[11pt]{article}
\pdfoutput=1
\usepackage{amssymb,amsmath,xcolor,empheq,physics,mathtools,url}

\numberwithin{equation}{section}
\usepackage{epsfig}
\usepackage{epstopdf}
\usepackage{latexsym}
\usepackage{graphicx}
\usepackage{booktabs}
\usepackage{bbm}
\usepackage{enumitem}
\usepackage[numbers,compress]{natbib}
\usepackage{bm}
\usepackage[T1]{fontenc}
\usepackage[linesnumbered,ruled]{algorithm2e}
\usepackage{amsthm,complexity,ytableau,youngtab,varwidth,calc}

\usepackage[margin=20pt,small]{caption}
\usepackage{subcaption}

\usepackage[toc]{appendix}

\usepackage[numbers,compress]{natbib}
\usepackage[ colorlinks=false, linkcolor=darkblue,
urlcolor=blue,
filecolor=blue,
citecolor=red, linktocpage=true, pdfstartview=FitV, bookmarksopen=true ]{hyperref} \hypersetup{pagebackref=true} \RequirePackage{doi}

\newcommand{\QXC}{\mathsf{QXC}}

\newcommand\equ[1] {\begin{equation}#1\end{equation}}

\newcommand\eqs[1] {\begin{align}#1\end{align}}
\newcommand\eqss[1] {\begin{align}\begin{split}#1\end{split}\end{align}} 

\renewcommand\( {\left(}
\renewcommand\) {\right)}

\usepackage[cmtip,all]{xy}
\newcommand{\longsquiggly}{\xymatrix{{}\ar@{~>}[r]&{}}}

\def\cV{\mathcal V}
\def\cH{\mathcal H}

\def\cO{\mathcal O}

\newtheorem{cor}{Corollary}

\newtheorem{thm}{Theorem}

\newtheorem{claim}{Claim}

\usepackage{tikz}
\usetikzlibrary{quantikz}
\usepackage{tikz-cd}

\newcounter{mycount}

\begin{document}  

\begin{titlepage}

\begin{center} 

\vspace*{1.5cm}

{\huge Quantum Spin Chains and Symmetric Functions  }

\vspace{1.5cm}

{\large   Marcos Crichigno and Anupam Prakash  
}

\bigskip
\bigskip
 QC Ware Corp., Palo Alto

\vskip 5mm
\today


\end{center}

\vspace{0.5cm}

\noindent 

\abstract{
We consider the question of what quantum spin chains naturally encode in their Hilbert space. It turns out that quantum spin chains are rather rich systems, naturally encoding solutions to various problems in combinatorics, group theory, and algebraic geometry. In the case of the XX Heisenberg spin chain these are given by skew Kostka numbers, skew characters of the symmetric group, and Littlewood-Richardson coefficients. As we show, this is revealed by a fermionic representation of the theory of “quantized” symmetric functions formulated by Fomin and Greene, which provides a powerful framework for constructing operators extracting this data from the Hilbert space of quantum spin chains. Furthermore, these operators are diagonalized by the Bethe basis of the quantum spin chain.  Underlying this is the fact that quantum spin chains are examples of  “quantum integrable systems.” This is somewhat analogous to bosons encoding permanents and fermions encoding determinants. This points towards considering quantum integrable systems, and the combinatorics associated with them, as potentially interesting targets for quantum computers.}

\end{titlepage}


\setcounter{tocdepth}{2}
\tableofcontents

\newpage


\section{Introduction and main results}

The central challenge in quantum computing is to identify computational problems that quantum mechanical systems solve more efficiently than classical systems. Much work has been devoted to this question since the landmark discovery by Shor that quantum mechanical systems can solve the integer factoring problem exponentially faster than any known classical method \cite{365700}.  Since then, enormous effort has been devoted to taking a computational problem of interest, developing quantum algorithms solving it, and investigating the nature of the quantum speedup, if any.\footnote{See, e.g., \cite{childs2010quantum,montanaro2016quantum} for a  survey of important quantum algorithms and the complexity zoo repository \href{}{https://quantumalgorithmzoo.org}.}

In this paper, we take a complementary view, which starts from considering quantum mechanical systems and searches for interesting computational problems arising there. This leads us to the question: {\it What do the simplest quantum mechanical systems naturally “want” to compute?} 
This has already been a fruitful line of reasoning; one may say that free bosons and fermions naturally want to compute permanents and determinants, which ultimately led Aaronson and Arkhipov  \cite{aaronson2010computational} to propose boson sampling as a basis for quantum supremacy demonstrations.

\begin{figure}[htbp]
\begin{center}
\includegraphics{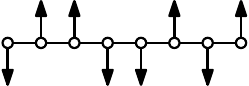}
\caption{A quantum spin chain. Each site consists of a quantum spin-$\tfrac12$ particle.}
\label{default}
\end{center}
\end{figure}

Here we consider the next simplest quantum many-body system in line: quantum spin chains.  These are one-dimensional systems of spin-$\tfrac12$ particles with nearest neighbor interactions, introduced by Heisenberg in 1928 as simple models for magnets. As we discuss, in a precise sense discussed below, we show that quantum spin chains  “want”  to compute solutions to a number of combinatorial, group theoretic, and geometric problems which are known to be hard to compute. 

The computational aspects of spin chains is less evident than that of the linear optics of bosons or fermions. This is revealed, however,  when exploiting the fact that Heisenberg spin chains are examples of  “quantum integrable systems.” As we review, these systems containing a large number of mutually commuting operators $\{\cO_I\}$ which can be explicitly constructed by the algebraic Bethe ansatz. This set includes the Hamiltonian and thus the $\{\cO_I\}$ generate  symmetries of the system. Since the product of symmetries is a symmetry, it follows that 
\equ{\label{cOintro}
\cO_I\cO_J = \sum_{K} C_{IJK}\cO_K \,,
}
where the $C_{IJK}$ are certain structure constants, which depend on the system at hand. It is in these structure constants that various computational problems are encoded. Although the algebra of symmetries above is rather general, we will mostly focus on the case of the XX spin chain, the simplest Heisenberg spin chain.

As we discuss, the theory of symmetric functions (i.e., the study of functions that are symmetric in a set of variables $x_i$) plays a crucial role in elucidating the computational problems encoded in XX quantum spin chains.  The theory of symmetric functions a rich field of mathematics with a number of applications in  combinatorics, group theory, Lie algebras, and algebraic geometry \cite{stanley1999enumerative,macdonald1998symmetric,egge2019introduction}. It can be extended to the case of functions depending on non-commuting variables, $\hat x_i$, which are often taken to satisfy some non-commutative algebra \cite{gelfand1994noncommutative,fomin1998noncommutative}. As we show, a fermionic representation of the formalism by Fomin and Greene \cite{fomin1998noncommutative} leads to an organizing principle for operations on the Hilbert space of the quantum spin chain. 

\

The two main results in the paper are the following. First, we show that a fermionic representation of the Fomin-Greene theory leads to a systematic method for turning symmetric functions $f(\mathbf{x})$ into operators $\hat f$ acting on the Hilbert space of quantum XX spin chains, which we refer to as ``quantized symmetric functions.'' Given a partition $\lambda\vdash d$, we identify three classes of quantized symmetric functions; $\hat h_\lambda$, $\hat s_\lambda$, and $\hat p_\lambda$, satisfying the algebra
\eqss{\label{prodshpsquantumintro}
\hat h_\mu \hat s_\nu =\,& \sum_{\lambda} K_{\lambda/\nu,\mu}\hat s_\lambda\,,\\
\hat p_\mu \hat s_\nu =\,& \sum_{\lambda} \chi^{\lambda/\nu}(\mu) \hat s_\lambda\,,\\
\hat s_\mu \hat s_\nu =\,& \sum_{\lambda} C_{\mu\nu}^{(q)\;\lambda} \hat s_\lambda\,.
}
The structure constants here correspond to (skew) Kostka numbers, (skew) characters of the symmetric group, and ($q$-deformed) Littlewood-Richardson coefficients, respectively, and encode solutions to combinatorial, group theoretic, and geometric problems \cite{stanley1999enumerative,macdonald1998symmetric,egge2019introduction}. 

Second, we show that the existence of the operators in \eqref{prodshpsquantumintro} is related to the quantum integrability of the underlying quantum spin chain. Indeed, all these operators are mutually commuting and are diagonalized in the Bethe basis of the XX spin chain. The Hamiltonian of the XX spin chain is given by one of the operators $\hat p_\mu$ and thus all these generate symmetries of the quantum spin chain. As we discuss in Section~\ref{sec:Diagonalization by Bethe ansatz} this matrix change of basis can be implemented efficiently by a quantum circuit. 

\

As we discuss below, one may act with with both sides of  \eqref{prodshpsquantumintro}  on a certain reference state and then the coefficients become amplitudes in a quantum state. Thus, one may say that the answer to  what XX quantum spins want to compute is skew  Kostka numbers, skew characters of the symmetric group, and Littlewood-Richardson coefficients.  Note that this is somewhat distinct from the sense in which free bosons and fermions want to compute permanents and determinants. In that case, it is the unitary evolution generated by a free Hamiltonian in a generic basis that leads to permanents and determinants appearing as amplitudes in the time-evolved quantum state. In contrast, the coefficients appearing in \eqref{prodshpsquantumintro} appear by the action of non-unitary operators and are discrete quantities, not parameterized by continuous variables. On the other hand, from the perspective of the quantum integrability of the spin chain, one may think of the operators in \eqref{prodshpsquantumintro} as interacting Hamiltonians.\footnote{More precisely, the hermitian Hamiltonians are given by the real and imaginary parts of these operators.} This suggests that implementing these Hamiltonians as quantum circuits (either as unitary time evolution or by embedding them in unitary matrices) may be useful in defining computational tasks associated to these coefficients. Although not the main focus here, we comment on this in Section~\ref{sec:Diagonalization by Bethe ansatz} and describe a unitary quantum circuit diagonalizing these operators. As we discuss in Section~\ref{sec:Discussion and open problems}, this raises the question of whether the more elusive Kronecker and plethysm coefficients can be understood in this setting. A quantum sampling algorithm for Kronecker coefficients was described in \cite{bravyi2023quantum}.

\ 

A few comments on computational complexity are in order.  The coefficients appearing on the RHS of \eqref{prodshpsquantumintro} are all at least $\#\P$-hard functions (see Appendix~\ref{app:Computational complexity} for an overview and references). Thus,  one does not expect an efficient algorithm, classical or quantum, computing them exactly. However, the complexity of approximation or sampling these quantities is less understood. The question of approximating characters of the symmetric group was considered by Jordan \cite{jordan2008fast}, where it was shown there that a classical algorithm can spoof a quantum algorithm approximating the {\it normalized} characters to additive accuracy. From the perspective of the methods used here, the characters of the symmetric group are rather simple, corresponding to a linear change of basis, while skew characters, skew Kostka numbers  and Littlewood-Richardson coefficients appear in the product ring \eqref{prodshpsquantumintro} and thus involve some nonlinearity.

The fact that a certain quantity appears naturally in a quantum mechanical system does not, by itself, automatically imply that it cannot be approximated or sampled from efficiently by classical means. Indeed, the determinants appearing in the wave functions for fermionic optics can be computed efficiently by classical methods and there are efficient classical algorithms for sampling from determinantal point processes. Potential quantum speedups related to the fermionic models must go beyond simply sampling from the wave function.  It is only for permanent {\it sampling} problems that bosonic optics is believed to provide an exponential speedup. In the same way, it is not immediately obvious whether the structure constants that quantum spin chains admit exponential speedups for sampling or additive approximation. Indeed, algorithmic techniques that go beyond sampling and amplitude estimation may be 
needed to obtain substantial speedups out of these systems. However, speedups are not out of the question either, especially for general quantum spin chains or other quantum integrable systems.  Indeed, although Heisenberg spin chains are amongst the simplest quantum mechanical systems, and there are powerful techniques to study them, classical computational methods can fall short. For example, placing the XXX spin chain in a random background magnetic field exhibits a transition between  thermal and localized phases which,  despite significant efforts, lies beyond classical computational methods. Based on this, spin chains on random background magnetic fields were proposed in 2018 by Childs et al. \cite{Childs_2018} as candidates for a demonstration of quantum supremacy. Although the problems we consider here are very different from that of \cite{Childs_2018}, their considerations still apply; these are simple quantum mechanical systems for which certain information lies beyond the grasp of classical methods. This raises the question of whether some of the mathematical quantities appearing in general Heisenberg spin chains admit a quantum speedup.

More broadly, one may propose a bottom-up approach to the search for good targets for quantum computers: identify the computational problems that naturally arise {\it within} quantum mechanical systems, starting from the simplest ones and working the way up towards more complex quantum systems, until quantum advantage is identified. The challenge in this approach is to identify the problems of potential use {\it outside} of quantum mechanics itself. In contrast, a top-down approach starts with a problem already known to be of interest outside of quantum mechanics, and the challenge is to determine whether the problem is well suited to quantum computers. The two approaches are complementary, although most efforts take the top-down approach.

\paragraph{Relation to previous work.} The relation between properties of quantum integrable systems, in particular spin chains, and various problems in enumerative geometry and quantum geometry is well known. It was observed by  Nekrasov and Shatashvili \cite{nekrasov2009supersymmetric,nekrasov2009quantum} that the quantum geometry of various algebraic varieties (precisely, cotangent bundles over  Grassmannian $G(k,n)$) is captured by the physics of quantum spin chains, in particular the XXZ quantum spin chain, which served as initial inspiration for this work. These ideas were further developed by Okounkov and collaborators \cite{maulik2012quantum}. The quantum geometry of the Grassmannian $G(k,n)$ was considered in \cite{GORBOUNOV2017282}, corresponding to the XX spin chain we consider here.  We note that by using  the Jordan-Wigner transform one can formulate these problems in terms of fermions and we find it convenient to do so. Fermions have been used to derive related results  by Paul Zinn-Justin and others; see  \cite{zinnjustin2009littlewoodrichardson} and  \cite{zinn2008integrability} for an overview. Here we shed new light on many of these results by introducing a new set of tools based on the (quantized) theory of symmetric functions \cite{gelfand1994noncommutative}, in particular the setting of Fomin and Greene \cite{fomin1998noncommutative}.\footnote{This is often referred to as the theory of symmetric functions in non-commuting variables or the theory of non-commuting symmetric functions. Here we consider a particular representation of the non-commuting variables in terms of operators in the Hilbert space of quantum spin chains and we refer to the functions thus obtained as  “quantized” symmetric functions. } This formalism allows us to derive in a simplified manner some of the existing results but also derive new results. Although we focus primarily on the calculus associated to the XX quantum spin chain here, we believe that the theory of quantized symmetric functions provides an organizing principle to attack various problems in enumerative combinatorics using quantum mechanical systems.  

\section{Heisenberg spin chains}

In this section, we give a brief overview of Heisenberg spin chains, the Bethe ansatz, and the Jordan-Wigner transformation. A spin chain is a 1d chain (open or closed) with $n$ sites, with a spin-$\tfrac12$ particle at each site. The Hilbert space is given by $\cH=\(\Bbb C^2\)^{\otimes n}$, where the basis for each factor is $(\ket{\uparrow},\ket{\downarrow})$. The  Heisenberg Hamiltonian is given by the simple 2-body interactions
\equ{\label{HXYZ}
H_{\text{XYZ}}= \sum_{i}\( J_{X} \, X_{i+1}X_i + J_Y \, Y_{i+1}Y_i + J_Z \, Z_{i+1}Z_i\)\,,
}
where $X,Y,Z$ are the standard Pauli matrices acting at site $i$ and the index runs over the sites of the chain (with its range depending on whether it is an open or closed chain) and the $J_{X,Y,Z}$ are real coupling constants. For arbitrary values of $J_{X,Y,Z}$ this is known as the XYZ  quantum spin chain.\footnote{One can also define Heisenberg models on general graphs, with interactions between sites connected by edges.} Special cases are:
\eqss{
\text{XX}:\,& \qquad J_X=J_Y\,,\qquad J_Z=0\,,\\
\text{XXX}:\,& \qquad J_X=J_Y=J_Z\\
\text{XXZ}:\,& \qquad J_X= J_Y\,,\qquad J_Z=\Delta\,,\\
\text{XYZ}:\,& \qquad J_X\neq J_Y\neq J_Z\,.\\
}
We will mostly focus here on the XX and XXZ quantum spin chains. Although simple systems, and despite efforts over nearly a century, the physics of Heisenberg spin chains is not full understood (see, e..g,  \cite{baxter2016exactly,eckle2019models}). 

\subsection{The coordinate Bethe ansatz and the Bethe equations}

In 1931 Hans Bethe proposed an ansatz for the eigensates of quantum spin chains above, now known as the Bethe ansatz. The original method was developed for the XXX spin chain but it can be extended to a general XYZ spin chain. Let us focus on the XXZ spin chain and set $J=1$.  Consider the magnetization sector $M=-\frac{k}{2}$, i.e., $k$ spins down and denote the positions of the down spins by $\{n_i,i=1,\ldots, k\}$ and the corresponding state by  $\ket{n_1,n_2,\ldots, n_k}$. This is referred to as the  “magnon” sector $k$. The Bethe ansatz postulates that all energy eigenstates in this sector take the form 
\equ{\label{Bethestatecoord}
\ket{\psi(p_1,\ldots,p_k) }=\sum_{1\leq n_1\leq \cdots \leq n_k\leq n} a({n_1,\ldots, n_k}) \ket{n_1,\ldots, n_k}\,,
}
with 
\equ{
a({n_1,\ldots, n_k}) =\sum_{\sigma \in S_k}  e^{ip_{\sigma_i}n_i+i \sum_{i<j} \theta_{\sigma_i\sigma_j}}\,,
}
where the sum is over all permutations and the parameters $\theta_{ij}\in [0,2\pi)$ define a  2-body scattering matrix $A_{ij}:=e^{i\theta_{ij}}$. The precise form of the scattering matrix $A_{ij}$ depends on the system at hand but are generally a function of the momenta $A_{ij}=A(p_i,p_j)$.  The Bethe equations are given by
\equ{\label{betheeqs}
e^{i p_i n}=\prod_{j\neq i} A(p_i,p_j)\,.
}
The solutions (for the momenta $p_i$) are known as Bethe roots. The interpretation of this equation is that when the $i$-th particle moves around the spin chain, it scatters through all other particles, picking up the corresponding scattering phases.  Thus, the many-body wavefunction is completely specified by the 2-body scattering. This remarkable feature can be seen as a defining feature of integrable models. Defining  $u_i = 2 \cot \frac{p_i}{2}$, the   scattering matrix for the XXZ spin chain is given by
\equ{
A(p_i,p_j)=\frac{u_i-u_j+i \Delta  }{u_i-u_j-i\Delta }\,.
}
Note that for the XX spin chain, $\Delta=0$, one has $A(p_i,p_j)=1$ so there are no scattering phases and the Bethe equations are simply
\equ{
e^{i p_i n}=1\,.
}
Thus, the $x_i:=e^{i p_i}$ are given by $k$-subset of the roots of unity. We define the set of Bethe roots as
\equ{\label{Betherootsunity1}
\cV_{k,n} = \{(x_1,\ldots, x_k)\; |\; x_i^n=1\,,  x_i\neq x_j\}/S_n\,.
}
Note that the $\text{dim}(\cV_{k,n} )={n \choose k}$ matches the dimension of the Hilbert space in this sector. In fact, one can show that the collection of  states \eqref{Bethestatecoord} evaluated at each Bethe root are  an orthogonal basis for the Hilbert space. 

\subsection{The Hilbert space and Young tableaux}
\label{sec:The Hilbert space and Young tableaux} 

\begin{figure}[t]
\begin{center}
\includegraphics{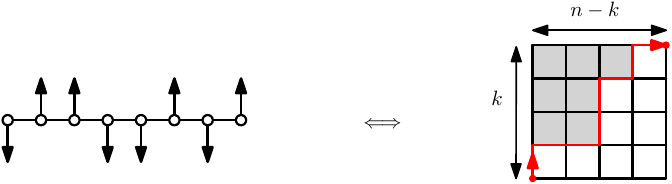}
\caption{A Young diagram $\lambda$ inside a $k\times (n-k)$ box  can be  represented by a path $\gamma(\lambda)$, from  the bottom left corner to the upper right corner of the rectangle. Such a path can be represented by an $n$-bit string $\gamma(\lambda)\in \{0,1\}^n$ of Hamming weight $k$, where a $0$ denotes a step to the right and a $1$ denotes a step up. This $n$-bit string also defines a classical configuration of spin-$\tfrac12$ chain with $n$ sites and magnon number $k$.}
\label{YTpath}
\end{center}
\end{figure}
Let us first set up a convenient notation for the basis states of the spin chain. Consider a spin chain of length $n$ in the magnetization sector $M=-\frac{k}{2}$, i.e, $k$ spins down. There are a total of $n\choose k$ states in this sector. To each such state one can associate a sequence of non-decreasing integers $\lambda= (\lambda_1,\ldots, \lambda_l)$, where  $l\leq k$ and $\lambda_1\leq n-k$. To see this, consider a box of size $k\times(n-k)$, containing the partition $\lambda$, as shown on the right of Figure~\ref{YTpath}. The partition can then be specified by the path delineating its boundary, starting from the bottom left corner to the upper right corner. Then, for each step north we associate a $\ket{\downarrow}$ and for each step to east  we associate a $\ket{\uparrow}$, leading to a spin chain basis with a total of $k$ spins down and  $n-k$ spins up. A distinguished state is given by the empty partition $\lambda = \emptyset$, corresponding to
\equ{
\ket{\emptyset}:= \ket{ \downarrow^k \, \uparrow^{n-k}}\,,
}
with $k$ consecutive  $\ket{\downarrow}$’s and  then  $(n-k)$ consecutive $\ket{\uparrow}$’s. We will take this as a natural “reference” state. In the language of many-body systems, this is the Hartree-Fock state.

Then, operations on the quantum Hilbert space of spin chains correspond to operations on Young tableaux. To understand exactly which operations one should perform to extract useful combinatorial and group theoretic information we will make use of the theory of (quantized) symmetric functions (see Sections~\ref{sec:The classical theory of symmetric functions} and \ref{sec:The quantized theory of symmetric functions}). 

\subsection{The Jordan-Wigner transformation}
It will be convenient to work in a second quantized formalism, using fermionic and annihilation and creation operators, satisfying the canonical  anticommutation relations
\equ{\label{cananticomm}
\{a_i, a_j^\dagger\}= \delta_{ij}\,.
}
The Jordan Wigner transformation corresponds to writing 
\equ{
a_i = \(\prod_{j=1}^{i-1} Z_i\) S^-_i\,,\qquad a_i^\dagger = \(\prod_{j=1}^{i-1} Z_i\) S^+_i\,,
}
where $S_{\pm}=\frac12\(X\pm i Y\)$. Under this map, spin up  corresponds to no fermion and spin down to one fermion:
\equ{
\ket{\uparrow}= \ket{0}\,,\qquad \ket{\downarrow}= \ket{1}\,.
}
Then, the XXZ spin chain Hamiltonian is given by (up to an overall normalization):
\equ{
H_{\text{XXZ}}= \sum_i\( a_{i+1}^\dagger a_{i} + a_{i}^\dagger a_{i+1} + \frac{\Delta}{2}(2n_i-1)(2n_{i+1}-1)\)\,.
}
Note that for $\Delta=0$ this  the free fermion Hamiltonian.\footnote{In the case of a general XYZ spin chain there are additional non-fermion number preserving terms.}

\section{The classical theory of symmetric functions}
\label{sec:The classical theory of symmetric functions}

In this section, we give an overview of the theory of symmetric functions. The theory of symmetric functions has a number of applications in  combinatorics, group theory, Lie algebras, and algebraic geometry \cite{stanley1999enumerative,macdonald1998symmetric,egge2019introduction}.  As we shall see in Section~\ref{sec:The quantized theory of symmetric functions} the quantized version of symmetric function theory provides an organizing principle for the appropriate quantum operators on quantum spin chains. 
%
%
Let $\mathbf{x}=(x_1,x_2,\ldots)$ be an infinite set of variables. A {\it homogeneous symmetric function of degree $d$} is a formal power series
\equ{
f(\mathbf{x})=\sum_\alpha c_\alpha x^\alpha\,,
}
that is symmetric under the exchange of any set of variables. See Appendix~\ref{App:Symmetric functions} for more details.


\subsection{The $(e,h,p,s)$ bases for symmetric functions}

There are various bases for symmetric functions, each with its own interesting properties. Four standard bases are the elementary basis, $\{e_\lambda\}$, the complete basis, $\{h_\lambda\}$, the power sum basis, $\{p_\lambda\}$, and the Schur basis, $\{s_\lambda\}$.   Let us  begin with the $(e,h,p)$ basis. For any integer $k\geq 1$ one defines, 
\eqs{
e_r :=\,&\sum _{1\leq j_{1}<j_{2}<\cdots <j_{r}}x_{j_{1}}\dotsm x_{j_{r}}\,,\\
h_r:=\,&\sum _{1\leq i_{1}\leq i_{2}\leq \cdots \leq i_{r}}x_{i_{1}}x_{i_{2}}\cdots x_{i_{r}}\,,\\
p_r:=\,&\sum_{1\leq i} x_i^r\,.
}
It is convenient to define $e_0=h_0=1$. Then, consider any $\lambda=(\lambda_1,\lambda_2,\cdots,\lambda_l)\vdash d$, and define
\eqs{
e_{\lambda}:=\,&e_{\lambda_1} e_{\lambda_2}\cdots  e_{\lambda_l} \,,\\
h_{\lambda}:=\,&h_{\lambda_1} h_{\lambda_2}\cdots  h_{\lambda_l} \,,\\
 p_\lambda :=\,& p_{\lambda_1}p_{\lambda_2}\cdots  p_{\lambda_l}\,.
}
The collection of symmetric functions $\{e_\lambda\}$ for all $\lambda \vdash d$ is a complete basis for symmetric functions of degree $d$. That is, any symmetric function of degree $d$ can be written as a a linear combination $f(\mathbf{x})=\sum_\lambda \alpha_\lambda e_\lambda(\mathbf{x})$, with the $\alpha_\lambda$ some coefficients. The sets $\{h_\lambda\}$ and $\{p_\lambda\}$ are also a complete basis. Perhaps the most important basis of all, however, is the Schur basis, defined as follows.  Given a partition $\lambda$, a semistandard Young tableaux (SSYT) of shape $\lambda$ is an array $T=(T_{ij})$ of positive integers of shape $\lambda$ that is weakly increasing in every row and strictly increasing in every column. Then, for a partition $\lambda = (\lambda_1, \lambda_2, \ldots, \lambda_n)$, the Schur functions is a sum of monomials,
\equ{
 s_\lambda(\mathbf{x})=\sum_{T\in \text{SSYT}(\lambda)} x^T  \,,
}
where the summation is over all SSYTs  $T$ of shape $\lambda$ and the exponents each $t_i$ counts the occurrences of the number $i$ in $T$. In other words, Schur polynomials are the generating functions of SSYTs. As an example, considering the SSYTs for the partition $\lambda=(2,1)$, one has
\equ{
s_{(2,1)}(\mathbf{x})= x_1^2x_2 + x_2^2x_1+x_1^2x_3+x_2^2x_3+\cdots
}
The Schur functions $\{s_\lambda\}$ for all $\lambda \vdash d$ are also a complete basis for the space of symmetric functions of degree $d$. 
If one is interested in properties of polynomials of a bounded degree, it is often sufficient to truncate the infinite set of variables $\mathbf{x}$ to a finite set by setting all variables $x_i=0$ for all $i> k$ and one defines
\equ{\label{truncsc}
s_\lambda(x_1,\ldots,x_k)=s_\lambda(x_1,\ldots,x_k,0,\ldots)\,.
}
This  statement will have a counterpart in the quantum setting which we use later. 

\subsection{Transition matrices and Hall inner product}
 A basic problem in the theory of symmetric functions is, given a symmetric function $f(\mathbf{x})$ in one basis, to find the coefficients in another basis. It turns out that these changes of basis encode solutions to combinatoric and group theoretic problems. Of particular interest to us here are the following relations:
\eqss{
e_\lambda=\sum_\mu K_{\lambda^T,\mu}s_\mu\,,\qquad 
h_\lambda=\sum_\mu K_{\lambda,\mu}s_\mu\,,\qquad 
p_\lambda=\sum_{\mu}\chi^{\mu}({\lambda})\, s_\mu\,,
}
where the $K_{\lambda,\mu}$ are known as skew Kostka numbers, the $\chi^{\mu}({\lambda})$  as skew characters of the symmetric group, and $\lambda^T$ stands for the transposed partitions. In the special case $\nu=\emptyset$ these are the standard Kostka numbers and characters. 

\

An important notion is that of the Hall inner product, denoted $\langle \cdot, \cdot \rangle$ (see Appendix~\ref{app:The Hall inner product} for definitions). The Schur basis is special in that it is the unique basis that is orthogonal (or self-dual) with respect to the Hall inner product:
\equ{\label{orthoschursec}
\langle s_\lambda, s_\mu\rangle  = \delta_{\lambda \mu}\,.
}
Then, the change of basis relations above can be written as
\equ{
\langle s_\nu,e_\mu\rangle=K_{\mu,\nu^T}\,,\quad
 \langle s_\nu,h_\mu\rangle=K_{\mu,\nu}\,,\quad 
 \langle s_\nu,p_\mu\rangle=\chi^{\nu}(\mu)\,.
}

\subsection{Generating functions}

Generating functions are important objects, collecting various symmetric functions into one object. Precisely, one introduces an auxiliary variable $t\in \Bbb C$ and then the generating functions for the $e_k,h_k$ and $p_k$ polynomials are defined as\footnote{We include a factor of $1/k$ in the definition of generating function for the power sum polynomials for later convenience.}
\equ{\label{defGeGhGp}
E(t):= \sum_{k\geq 0} t^k\, e_k\,,\qquad H(t):= \sum_{k\geq 0} t^k\, h_k\,,\qquad \Psi(t):=\sum_{k=1}^{\infty} \frac{t^k}{k}  \, p_k \,.
}
This should be thought as  formal power series in $\mathbf{x}$ and $t$.   Note we omit the dependence on $x$ to simplify notation. It is easy to see that the $e$- and $h$-generating functions  can be written compactly as 
\equ{
E(t)=\prod_{i\geq 1}(1+tx_i)\,,\qquad H(t)=\prod_{i\geq 1}\frac{1}{1-tx_i}\,.
}
That is, expanding the RHS of this expression in powers of $t$, each term in the (infinite) expansion matches the terms in \eqref{defGeGhGp} at each power of $t$, as can be easily checked. Note that 
\equ{
E(t)H(-t)=1\,.
}
Thus, the elementary and complete symmetric functions are “inverses” of each other in this sense.  Furthermore, one can show that the generating function for the complete and elementary symmetric polynomials is the exponential of the generating function for the power sum polynomials:
\equ{\label{classicaHEP}
H(t)=e^{\Psi(t)}\,,\qquad E(t)=e^{-\Psi(-t)}\,. 
}
One can show that these follow from the Girard–Newton formulae relations relating the power sum functions to the elementary and complete symmetric functions going all the way back to the 1600s. We will see later that there is a quantum version of these relations. 

\

A useful set of identities are known as the Cauchy identities  \cite{stanley1999enumerative}:
\eqss{\label{classicalcauchy}
\sum_{\lambda} s_\lambda (x)s_\lambda(y)=\prod_{i,j}\frac{1}{1-x_i y_j}\,,\qquad 
\sum_{\lambda} s_\lambda (x)s_{\lambda^T}(y)=\prod_{i,j}(1+x_i y_j)\,.
}
There will be a quantum counterpart of these equations.

\subsection{The ring of symmetric functions}

The product of two symmetric functions is a symmetric function. Thus, symmetric functions are endowed with a ring structure with the regular addition and multiplication. In particular, this means that the product of any two basis elements can be expanded in any other basis. Of special importance to us are:
\eqs{\label{prodshps}
h_\mu s_\nu =\,& \sum_{\lambda} K_{\lambda/\nu,\mu}s_\lambda\,,\\
p_\mu s_\nu =\,& \sum_{\lambda} \chi^{\lambda/\nu}(\mu) s_\lambda\,,\\
s_\mu s_\nu =\,& \sum_{\lambda} C_{\mu\nu}^{\;\;\;\;\lambda} s_\lambda\,.
}
We have already mentioned the coefficients appearing in the first two lines. The  coefficients $C_{\mu\nu}^{\;\;\;\;\lambda} $ are known as  the Littlewood-Richardson coefficients. These can be defined simply via this expression, i.e., as the structure constants in the multiplication ring of Schur functions. However, they can be also independently defined as the Clebsch-Gordan coefficients in the representation ring of $GL(n,\Bbb C)$ (see, e.g., \cite{sagan2013symmetric}). The Littlewood–Richardson rule is a combinatorial rule for computing these coefficients. Namely, as the the  number of “Littlewood–Richardson tableaux” of shape $\lambda/\mu$ and weight $\nu$ \cite{stanley1999enumerative}.\footnote{A Littlewood–Richardson tableau is a skew semistandard tableau whose reverse-row-concatenated sequence forms a lattice word, ensuring every $i$ appears no less than $i+1$ in any sequence segment.}
Note that in all these formulas $\abs{\lambda}=\abs{\mu}+\abs{\nu}$ for the degrees to match on both sides.  We will return to these formulas in the context of quantum spin chains in Sections~\ref{sec:The quantized theory of symmetric functions} and \ref{sec:The calculus of the XX quantum spin chain}.

\section{The quantized theory of symmetric functions}
\label{sec:The quantized theory of symmetric functions}

In this section, we discuss the theory of symmetric functions in non-commuting variables \cite{gelfand1994noncommutative}. We  consider in particular the theory developed by Fomin and Greene \cite{fomin1998noncommutative}.

\subsection{The Fomin-Greene formalism}

The general idea in the theory of symmetric functions in non-commuting variables is to promote the  variables $x_i$ to non-commuting variables
\eqss{
x_i \quad \to& \quad \hat x_i\,, \qquad \qquad [\hat x_i,\hat x_j]\neq 0\,.
}
Correspondingly, a function $f(x_i)$ is promoted to a corresponding $\hat f= f(\hat x_i)$.\footnote{Note that this is sometimes referred to as the theory of non-commuting symmetric functions but this terminology would lead to confusion in our setting as the functions of interest to us, although they depend on non-commuting variables, they are in fact commuting. }  As in any quantization, there will be ordering ambiguities in the definition of $\hat f$. In the formalism of Fomin and Greene \cite{fomin1998noncommutative}  the $\hat x_i$ are required to satisfy some “non-local” commutativity:
\eqs{\label{nonlocalcomm}
\hat x_i \hat x_j =\,&\hat x_j \hat x_i\,,\qquad \qquad  \abs{i-j}\geq 2\,,\\ \vspace{2mm}
\label{localcomm}
(\hat x_i +\hat x_{i+1})\hat x_{i}\hat x_{i+1}=\,&\hat x_{i}\hat x_{i+1}(\hat x_i +\hat x_{i+1})\,.
}
As discussed in \cite{fomin1998noncommutative}, this contains several known algebras as special cases.

\subsubsection{The $e,h$ and $s$ operators}

Elementary and complete homogeneous symmetric functions in the non-commuting variables $\hat x_i$  are  defined as (here we follow the quantization prescription in \cite{fomin1998noncommutative}):
\equ{\label{ekquantum}
\hat e_k := \sum_{ j_1 < j_2 < \cdots < j_k}\hat x_{j_{1}}\hat x_{j_{2}}\dotsm \hat x_{j_{k}}\,,\qquad 1\leq k \leq n\,.
}
\equ{\label{hkquantum}
\hat h_k:=  \sum _{ i_1\geq i_2 \geq \cdots \geq  i_k}\hat x_{i_{1}}\hat x_{i_{2}}\cdots \hat x_{i_{k}}, \qquad 1\leq k \leq n\,.
}
Note that an ordering for the $\hat x_i$ has been chosen. For the $\hat e_k$ they are increasing to the right and for the $\hat h_k$ they are decreasing to the right.\footnote{Note this is reversed with respect to the definitions in  \cite{fomin1998noncommutative}. Correspondingly, the order $\hat x_{i}\hat x_{i+1}$ in the condition \eqref{localcomm} is  reversed.} 
Note that \eqref{localcomm} is the condition that $\hat e_1(\hat x_i, \hat x_{i+1})$ and $\hat e_2(\hat x_i, \hat x_{i+1})$ commute. It is then shown in \cite{fomin1998noncommutative} that in fact  \eqref{nonlocalcomm} and \eqref{localcomm} imply that {\it all}  polynomials $\hat e_j$ commute, i.e., 
\equ{\label{commehat}
\hat e_j\hat e_k =\hat e_k  \hat e_j\,.
}
The  quantized Schur functions are defined as in the classical case in terms of SSYTY.  Given a partition $\lambda$, define the quantized Schur function $\hat s_\lambda(\hat x)$ by
\equ{\label{defShat}
\hat s_\lambda(\hat x) := \sum_T \hat x^T\,,
}
where the  sum ranges over all semi-standard tableaux $T$ of shape $\lambda$ and $\hat x^T$ denotes the product $\prod_i \hat x_{i}$, with indices $i$ obtained by reading the each column from the top-down, starting with the last column.\footnote{Again, this ordering convention is opposite to that of \cite{fomin1998noncommutative}. } 
For instance, 
\eqss{
 \hat s_{\tiny \yng(1)} =\,& \hat x_1 + \hat x_2+  \hat x_3+\cdots\,,\\
 \hat s_{\tiny \yng(1,1)}=\,& \hat x_1 \hat x_2+\hat x_1 \hat x_3 +\hat x_2  \hat x_3+\cdots\,,\\
\hat s_{\tiny \yng(2)} =\,& \hat x_1^2+\hat x_2^2+\hat x_3^2+\hat x_3 \hat x_2+\hat x_3 \hat x_1 +\hat x_2  \hat x_1+\cdots\,, \\
\hat s_{\tiny \yng(2,1)} =\,& \hat x_2\hat x_1^2+\hat x_2 \hat x_1 \hat x_2+\hat x_2 \hat x_1 \hat x_3+ \hat x_3 \hat x_1^2+\hat x_3 \hat x_1 \hat x_2+\hat x_3 \hat x_1 \hat x_3+\hat x_3\hat x_2^2+\hat x_3 \hat x_2 \hat x_3+\cdots\,,\\
 \hat s_{\tiny \yng(1,1,1)} =\,& \hat x_1  \hat x_2 \hat x_3+\cdots\,.
}
Note that with these definitions, 
\equ{
\hat s_k = \hat h_k\,,\qquad \hat s_{1^k} = \hat e_k\,, 
}
as in the case of commuting variables.  In fact,  one of the central results in \cite{fomin1998noncommutative} is that $\hat s_\mu$  satisfy a non-commutative version of the Jacobi-Trudi formula (Lemma 3.2): 
\equ{\label{JTop}
\hat s_{\lambda/\mu} = \det \(\hat h_{\lambda_i-\mu_j+j-i}\)=\det \(\hat e_{\lambda^T_i-\mu^T_j+j-i}\)\,,
}
where by definition $\hat e_0=1$ and $\hat e_k=0$ for $k<0$. Since the $\hat e_i$ commute with one another, there is no ordering ambiguity in this expression.
A direct consequence is that the quantized Schur functions commute with $\hat E(t)$, i.e., 
\equ{
[\hat s_\mu,\hat E(t)]=0\,,
}
which is a direct consequence that, under the assumptions \eqref{nonlocalcomm} and \eqref{localcomm}, the $\hat e_k$ commute.

 More generally,  since the $\hat e_k$ commute  any function of the $\hat e$ behaves exactly as in the case of commuting variables. As a consequence, the Schur functions satisfy all the regular properties of Schur functions. In particular, the $\hat s_\mu$ commute and their product expands just like ordinary Schur functions:
\eqs{\label{hatSSS}
 \hat s_{\mu}  \hat s_{\nu}=\,& \sum_{\lambda} C_{\mu \nu}^{\;\;\;\;\lambda}\hat s_\lambda\,, 
}
Note these are formally identical to the expressions for their commutative versions. Indeed, as discussed in \cite{fomin1998noncommutative}, whenever conditions \eqref{nonlocalcomm} and \eqref{localcomm} hold, all the identities of the commutative theory which can be expressed solely in terms of the $s_\lambda(x)$ hold for the non-commutative versions $\hat s_\lambda(\hat x)$. One can check this explicitly for the operators above. Then, using the algebra  \eqref{nonlocalcomm} and \eqref{localcomm} one can explicitly check that \eqref{hatSSS}  holds.

\subsubsection{Generating functions and Cauchy identities}

Now, consider the generating function $E(t)$ for the elementary symmetric polynomials. We define its quantized version by replacing $e_k\to \hat e_k$:
\equ{\label{defDop}
E(t)\to\hat E(t):= \sum_{i=0}^{\infty} \hat e_k t^k=\prod_{i=1}^{\infty}(1+t\hat x_i)\,.
}
where the order in the product is as shown.  Similarly, for the complete symmetric polynomials we define the quantized generating function as:
\equ{\label{defbarDop}
H(t)\to \hat H(t):= \sum_{i=0}^{\infty} \hat h_k t^k=\prod_{i=\infty}^{1}\frac{1}{1-t\hat x_i}\,,
}
Note that 
\equ{\label{hatEH1}
\hat E(t) \hat H(-t)=1\,,
}
which is the operator analog of $E(t) H(-t)=1$. Note that since  all $\hat e_k$ commute with each other, the corresponding generating functions also commute:
\equ{
[\hat E(t),\hat E(t’)]=[\hat H(t),\hat H(t’)]=[\hat E(t),\hat H(t’)]=0\,.
}
The classical Cauchy identities \eqref{classicalcauchy} have non-commutative analogs \cite{fomin1998noncommutative}.  We denote regular commuting variables by $t_i$ and non-commuting ones by $\hat x_i$.  Then, 
\eqs{
 \sum_{\lambda}  s_\lambda( t)  \, \hat s_{\lambda}(\hat y) =\prod_{i=1}^m \hat H(t_i)\,,  \label{NC-Cauchy}
}
Note that since the generating functions commute with one another, the ordering on the RHS is unimportant, consistent with the fact that in the LHS the functions are symmetric in the variables $t$. 

\subsection{The calculus of the XX quantum spin chain}
\label{sec:The calculus of the XX quantum spin chain}

In this section, we use a fermionic representation of algebra  \eqref{nonlocalcomm} and \eqref{localcomm}, as operators acting on fermionic Fock space. This allows us to identify the operations in the Hilbert space that extract the characters of the symmetric group,  Kostka numbers, and  Littlewood-Richardson coefficients by the action of fermionic operators in Fock space. Consider a (semi-infinite) set of fermionic modes satisfying the canonical anti-commutation relations \eqref{cananticomm} and let 
\equ{\label{defxi}
\hat x_i= a_{i+1}^\dagger a_i\,,
}
for all $i\geq 1$, be the operators that move a fermion from site $i$ to $i+1$ in a semi-infinite chain. Note these are infinite-dimensional operators as we consider a semi-infinite chain. It is easy to see that these  satisfy the conditions \eqref{nonlocalcomm} and \eqref{localcomm} and are thus a valid representation of the algebra and the Fomin-Greene operators can be written in terms of these fermionic operators. Note that these are infinite-dimensional operators as there are an infinite number of operators. However, we will see below that just as in the theory of classical symmetric functions, it is consistent to set all $\hat x_i$ for all $i>k$.   Note that the shift operators are nilpotent, 
\equ{\label{nilpotx}
\hat x_i^2=0\,.
}
Then, using this  \eqref{defbarDop} the generating function is given by
\equ{
\hat H(t)=\prod_{i=n-1}^{1}(1+t\hat x_i)\,, \qquad \hat E(t)= \prod_{i=1}^{n-1}(1+t\hat x_i)\,.
}
Note the only difference is the order in the product.  Expanding in powers of $t$ one can check that these coincide with the \eqref{ekquantum} and \eqref{hkquantum}. Note these operators also shift fermions $k$ units  but only if the intermediate sites are unoccupied. We give explicit expressions below. 

\

It turns out that $\hat H(t)$ can be written in an exponentiated form as
\equ{\label{Hvertex}
\hat H(t)=e^{\sum_{k=1}^{\infty} \frac{t^k}{k} \hat J_k}\,,
}
where the $J_k$ are  the “current operators” 
\equ{
\hat J_k = \sum_{i=1}^{\infty} a_{i+k}^\dagger a_i\,.
}
Note these operators shift fermions $k$ units to the right by hopping over any intermediate fermions, and picking up the corresponding signs.  It is easy to see that the currents commute with one another, 
\equ{
[\hat J_k,\hat J_{k’}]=0\,.
}
The exponential of currents in \eqref{Hvertex} is sometimes referred to as a “vertex operator.”  Note that comparing \eqref{Hvertex} to the classical relation \eqref{classicaHEP} suggests that the current operators are the quantization of the power sum polynomials and we write 
\equ{
J_k =\hat p_k\,.
}
We shall see below that this is indeed correct, by verifying that the operators $\hat p_\lambda$ satisfy all the expected relations in the multiplication ring of (quantized) symmetric functions. 

Finally, the quantized Schur functions $\hat s_\lambda$ are obtained by the setting \eqref{defxi} in the definition \eqref{defShat} or the quantized Jacobi-Trudy formulas \eqref{JTop}.

\

We now arrive at the main statement in this section. Since the operators $\hat h_\lambda$, $\hat p_\lambda$, and $\hat s_\lambda$ all commute with one another, any classical function involving these functions also holds as an operator equation. In particular, the  relations \eqref{prodshps} in the classical product ring symmetric functions become the operator relations 
\eqs{\label{prodshpsquantum}
\hat h_\mu \hat s_\nu =\,& \sum_{\lambda} K_{\lambda/\nu,\mu}\hat s_\lambda\,,\\
\hat p_\mu \hat s_\nu =\,& \sum_{\lambda} \chi^{\lambda/\nu}(\mu) \hat s_\lambda\,,\\
\hat s_\mu \hat s_\nu =\,& \sum_{\lambda} C_{\mu\nu}^{\;\;\;\;\lambda} \hat s_\lambda\,.
}
Recall that these are all formally infinite-dimensional operators, with action on the semi-infinite line. Note, however, that these functions have a finite degree $\abs{\mu}$ in the variables $\hat x_i$. Then, if acting on a state with a finite number of fermions, it can at most move the fermions by the finite amount $\abs{\mu}$ and there will never be any fermions present at some large enough site number $n$. Then,  the infinite expansions in \eqref{defShat}  truncate to a finite number of terms, effectively setting
\equ{\label{xiopen}
\hat x_i\to 0\qquad \qquad \text{for all $i\geq n$}\,.
}
Thus, we can consider truncated quantized Schur functions, as in the classical case \eqref{truncsc}. We can think of this as  considering an open chain of length $n$ (we will consider the case of a closed chain below). Consider now the  sector of fermion number $k\leq n$ on this open chain and the action of $\hat s_\mu$ on the empty partition  $\ket{\emptyset}=\ket{1^k0^{n-k}}$. It is easy to see by direct inspection that only one term in \eqref{defShat} acts nontrivially and the action of the quantized Schur function is simply to create the classical state corresponding to the partition $\mu$, i.e., 
\equ{\label{sstatemu}
\hat s_\mu \ket{\emptyset}= \ket{\mu}\,,
}
where the state $\ket{\mu}$ on the RHS is precisely the staircase representation of the partition $\mu$ described in Section~\ref{sec:The Hilbert space and Young tableaux}. That is, the computational basis of the spin chain correspond to the Schur basis. 

Now, recall that thee partitions $\mu$ arising here have at most $k$ rows and $n-k$ columns and thus fit inside a box of size $k\times(n-k)$. Thus,  if one is interested in properties of partitions $\mu$ of size up to $d$, then one must set $k\geq d$ and $n\geq d$.  Then, acting with the equations \eqref{prodshpsquantum}  on the Hartree-Fock state we obtain 
\eqs{\label{hnuK}
\hat h_\mu \ket{\nu} =\,& \sum_{\lambda} K_{\lambda/\nu,\mu} \ket{\lambda}\,,\\ \label{pnuchi}
\hat p_\mu \ket{\nu} =\,& \sum_{\lambda} \chi^{\lambda/\nu}(\mu) \ket{\lambda}\,,\\ \label{snuC}
\hat s_\mu \ket{\nu} =\,& \sum_{\lambda} C_{\mu\nu}^{\;\;\;\;\lambda} \ket{\lambda}\,.
}
Thus, we have shown that the action of the operators $\hat h_\mu \ket{\nu}$, $\hat p_\mu \ket{\nu} $, and $\hat s_\mu \ket{\nu} $ on the Hilbert space of the spin chain encode Kostka, characters of the symmetric group,  and Littlewood-Richardson coefficients respectively. 

Note that we have thus far considered the open chain. It is possible, however, to consider a closed chain with $n$ sites with twisted boundary conditions with parameter $q$. This corresponds to modifying \eqref{xiopen} by setting  
\equ{
\hat x_n=q a_1^\dagger a_n \,,\qquad \qquad \text{$x_i \to 0$ for all $i\geq n+1$}\,.
}
 As we shall see in more detail below, this leads to computing $q$-deformed $q$-deformed Littlewood-Richardson coefficients:
\equ{\label{ssCq}
\hat s_\mu\hat s_\nu =\sum_{\lambda}C_{\mu\nu}^{(q)} \hat s_\lambda\,.
}
These are a generalization of the standard ($q=0$) Littlewood-Richardson coefficients.\footnote{ Note that if one is interested in standard Littlewood-Richardson coefficients one can simply set $q=0$ here. Otherwise, one can always embed the partitions in a representation with a number of trailing zeros, $(\ket{\mu},\ket{\nu},\ket{\lambda})\to( \ket{\mu,0^m},\ket{\nu,0^m},\ket{\lambda,0^m}))$ with $m> \abs{\mu}+\abs{\nu}$ and then the coefficients generated above for these partitions are the  ($q=0$) Littlewood-Richardson coefficients.\footnote{In other words, with such a large number of trailing zero’s the operators will never get to see the $q$ sitting at the end of the chain of length $L=n+m$.} }

\subsection{Characters of the symmetric group and Kostka numbers}

We can provide another understanding on why the formulas above work. To see this, let us work out some examples explicitly. 

\paragraph{Characters.} 
Let us begin with the equation \eqref{pnuchi} and for simplicity we set  $\nu=0$, which corresponds to the standard characters of the symmetric group:
\equ{\label{pmucharh}
\hat p_\mu \ket{\emptyset} = \sum_{\lambda} \chi^{\lambda}(\mu) \ket{\lambda}\,.
}
Recall  the irreducible representations (irreps) of  the symmetric group $S_d$,  as well as the conjugacy classes, are  labeled by partitions $\lambda \vdash d$. The characters can be computed by explicitly constructing the corresponding matrix representations and taking the trace.\footnote{Recall that given an irrep $W_\lambda: S_d\to GL(V)$, the character of the representation on the conjugacy class $\mu$ is defined as $\chi^{\lambda}(\mu) = \text{Tr}\, W_\lambda(\mu)$.
}
Alternatively, a well known combinatorial rule is the  Murnughan-Nakayama rule, which states that 
\equ{
\chi^{\lambda}(\alpha)=\sum_T (-1)^{\text{ht}(T)}\,,
}
where the sum is over all border-strip tableaux $T$ of shape $\lambda$ and type $\alpha$.  It is this expression that explains why the fermionic operators $\hat p_\lambda$ capture the characters of the symmetric group. Note that in order to capture all the  characters we must set $k\geq d$ and $n\geq 2d$, the minimal choice being a chain of length $n=2d$ at half-filling $k=d$.

\begin{figure}[htbp]
\begin{center}
\includegraphics[scale=0.8]{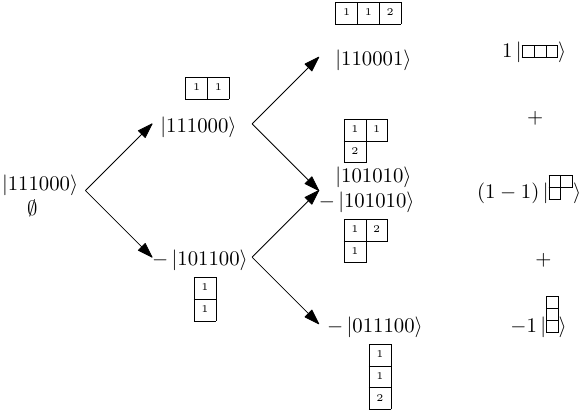}
\caption{Action on the empty partition $\ket{\emptyset}$ of the operator $\hat J_{\{2,1\}}=\hat J_1 \hat J_2$, corresponding to type $\alpha=\{2,1\}$ (i.e. filling with the numbers $\{1,1,2\}$). In the first step, $\hat J_2$ first creates two border strip partitions filled with the 1’s. In the second step, the action of $\hat J_1$ adds border strips with $2$’s. Note at this level there is a cancellation of the two tableaux ${\tiny \begin{ytableau}
1 &1 \\
2
\end{ytableau}}$ and ${\tiny \begin{ytableau}
1 &2 \\
1
\end{ytableau}}$  and the final quantum state $\hat J_{\{2,1\}}\ket{\emptyset}$ contains exactly the characters $\chi^\mu(\{2,1\})$. The rule for filling in the tableaux is if the $k$th fermion from the right moves this creates boxes in the $k$th row. This automatically implements the sign $(-1)^{\text{ht}(T)}$ from the Murnaghan-Nakayama rule and thus leads to the characters of the symmetric group. 
}
\label{fig:charS3fermionspath}
\end{center}
\end{figure}

To see this at work explicitly, consider $S_3$ and thus a chain of  length $n=6$ and fermion number $k=3$. The partitions are $(3), (2,1)$, and $(1^3)$. Using the Murnughan-Nakayama rule one obtains the characters shown in Table~\ref{table:charactersS3}.  Now, let us consider the states of the spin chain. There are in total ${6\choose 3}=20$ states. These include
\equ{
\ket{\emptyset}=\ket{111000}\,,\qquad \ket{\tiny \yng(3)}=\ket{110001}\,,\qquad \ket{\tiny \yng(2,1)}=\ket{101010}\,,\qquad \ket{\tiny \yng(1,1,1)}=\ket{011100}\,,
}
corresponding to the empty partition and the three partitions of $d=3$.\footnote{The other 14 other states in the Hilbert space are not relevant to this computation but do appear when computing other coefficients, as we shall see.}  The relevant operators are
\equ{
\hat J_{\tiny \yng(3)}=\hat J_{3}\,,\qquad \hat J_{\tiny \yng(2,1)}=\hat J_{2}\hat J_{1}\,,\qquad \hat J_{\tiny \yng(1,1,1)}=\hat J_{1}^3\,.
}
Acting with these explicitly one gets, 
\eqss{
\hat J_{\tiny \yng(3)}\ket{\emptyset}=\,&\ket{110001}-\ket{101010}+\ket{011100}=\ket{\tiny \yng(3)}- \ket{\tiny \yng(2,1)}+\ket{\tiny \yng(1,1,1)}\,,\\
\hat J_{\tiny \yng(2,1)}\ket{\emptyset}=&\hat p_2\ket{110100} =\ket{110001}-\ket{011100}=\ket{\tiny \yng(3)} -\ket{\tiny \yng(1,1,1)} \,,\\
\hat J_{\tiny \yng(1,1,1)}\ket{\emptyset}=&\,\hat p_1^2 \ket{110100}=\hat p_1\(\ket{110010}+\ket{101100}\) =\ket{\tiny \yng(3)}+2\ket{\tiny \yng(2,1)}+\ket{\tiny \yng(1,1,1)}\,.
}
We see these exactly reproduce the coefficients in  Table~\ref{table:charactersS3}. 
\begin{table}[]
\begin{center}
\begin{tabular}{|c|c|c|c|} \hline
 & $(3)$ & $(2,1)$ &$(1^3)$ \\ \hline
$\chi^{3}$ & $1$& $1$& $1$\\ \hline
$\chi^{2,1}$ &$-1$ &$0$ &$2$ \\ \hline
$\chi^{1^3}$ &$1$ &$-1$ & $1$ \\ \hline
\end{tabular}
\end{center}
\caption{Character table for $S_3$. See e.g. book by G.D. James, pg. 133. }
\label{table:charactersS3}
\end{table}%
Note that the sum in the RHS of \eqref{pmucharh} contains only states $\ket{\mu}$ with $\mu\vdash d$. It is not obvious that this should be the case; fermion number conservation property of $\hat J_\lambda $ only ensures that the partitions appearing on the RHS fit inside a $d\times d$ box, but not that they exactly contain $d$ boxes. The fact this is nonetheless true is a nontrivial fact.  A way to understand this is via the Murnughan-Nakayama rule. Indeed, it is easy to see that the staircase encoding of partitions with fermions, the operators $\hat J_\mu$ exactly implement the Murnughan-Nakayama rule, with the signs in the RHS coming from the signs due to fermions hopping over each other. The case of $S_3$  is shown in Figure~\ref{fig:charS3fermionspath}.

In fact, one can more generally define the skew characters $\chi^{\lambda/\mu}(\alpha)$ of the symmetric group and the Murnughan-Nakayama rule states that 
\equ{
\chi^{\lambda/\mu}(\alpha)=\sum_T (-1)^{\text{ht}(T)}
}
where the sum is over all border-strip tableaux $T$ of shape $\lambda/\mu$ and type $\alpha$. One can check that this is correctly reproduced by the general formula \eqref{pnuchi}. 

\paragraph{Kostka numbers.}

The Kostka numbers $K_{\mu, \lambda}$ are defined as the number of semistandard Young tableaux of shape $\mu$ and weight $\lambda$. Thus, one can write
\equ{
K_{\mu, \lambda}=\sum_T 1\,,
}
where the sum is over  semistandard Young tableaux of shape $\mu$ and weight $\lambda$.  Note the only difference with the Murnughan-Nakayama rule for the characters is the absence of minus signs. 
\begin{table}[t]
\begin{center}
\begin{tabular}{|c|c|c|c|} \hline
 & $(3)$ & $(2,1)$ &$(1^3)$ \\ \hline
$K_{3}$ & $1$& $1$& $1$\\ \hline
$K_{2,1}$ &$0$ &$1$ &$2$ \\ \hline
$K_{1^3}$ &$0$ &$0$ & $1$ \\ \hline
\end{tabular}
\end{center}
\caption{Kostka numbers for all partitions of size exactly 3, taken from Wikipedia’s article on Kostka numbers. }
\label{table:Kostka3}
\end{table}%
To see how this is implemented by the operators $\hat h_\lambda$, take as an example again the partitions $\lambda\vdash 3$. It is easy to directly count the number of SSYTs of different types explicitly and the resulting Kostka numbers are shown in  Table~\ref{table:Kostka3}. On the other hand, acting with the corresponding operators on the Hartree-Fock state gives
\eqs{
\hat h_{3}\ket{111000} =\,& \ket{110001}=\ket{\tiny \yng(1,1,1)}\,,\\
\hat h_{2,1}\ket{111000} =\,&\ket{101010}+\ket{110001}=\ket{\tiny \yng(3)}+\ket{\tiny \yng(2,1)}\,,\\
\hat h_{1^3}\ket{111000} =\,& \ket{011100}+2\ket{101010}+\ket{110001} =\ket{\tiny \yng(1,1,1)}+2\ket{\tiny \yng(2,1)}+\ket{\tiny \yng(3)} \,,
}
reproducing these numbers.  Note that, in contrast to the calculation of characters, no negative numbers can appear here, since the fermions never hop over each other and thus their anticommuting nature is not revealed. This is consistent with the Kostka numbers being non-negative. 

\subsection{Littlewood Richardson coefficients}

Let us consider a chain of length  $n$ with fermion number $k$. We allow for a closed chain with twisted boundary conditions set by a parameter $q$. That is, 
\equ{\label{repxLRq}
\hat x_i= a_{i+1}^\dagger a_i \quad \text{for $i<n$}\,,\qquad \text{and} \quad \hat x_n= q a_{1}^\dagger a_n \,.
}
The meaning is that if a fermion goes around the chain site $n$ back to site $1$ it picks up a factor of $q$. Setting $q=0$ corresponds to an open chain and $q=\pm 1$ to antiperiodic and periodic boundary conditions. Note that \eqref{sstatemu} still holds for arbitrary $q$ since the term $a_{1}^\dagger a_n$ vanishes on $\ket{\emptyset}$. To see that this indeed leads to  $q$-deformed Littlewood-Richardson coefficients \eqref{ssCq}, we consider the example $n=4$ and $k=2$. The quantized Schur polynomials are obtained from setting $n=4$ in  \eqref{repxLRq} and replacing this in  \eqref{defShat}. For the for single-row partitions one finds:
\eqss{
 \hat s_{\tiny \yng(1)} =\,& a_2^\dagger a_1 + a_3^\dagger a_2 +a_4^\dagger a_3 + q a_4^\dagger a_1 \,,\\
 \hat s_{\tiny \yng(2)} =\,& a_3^\dagger \nu_2 a_1 +a_4^\dagger \nu_3 a_2+ a_4^\dagger a_3 a_2^\dagger a_1+q  \(a_4\nu_1 a_2^\dagger +\nu_4 a_3 a_1^\dagger+ a_4 a_3^\dagger a_2 a_1^\dagger\)\,,\\
 \hat s_{\tiny \yng(3)} =\,& a_4^\dagger \nu_3 \nu_2 a_1+q \(\nu_4\nu_3a_2a_1^\dagger +a_4 a_3^\dagger\nu_2 \nu_1 +\nu_4 a_3 \nu_1 a_2^\dagger\)\,,\\
 \hat s_{\tiny \yng(4)} =\,& q \nu_4 \nu_3 \nu_2 \nu_1\,,
}
where $\nu_i=1-n_i$ with $n_i$ the number operator. For single-column partitions
\eqss{
 \hat s_{\tiny \yng(1)} =\,& a_2^\dagger a_1 + a_3^\dagger a_2 +a_4^\dagger a_3 + q a_4^\dagger a_1 \,,\\
 \hat s_{\tiny \yng(1,1)} =\,& a_3^\dagger n_2 a_1 +a_4^\dagger n_3 a_2+ a_4^\dagger a_3 a_2^\dagger a_1+q  \(a_4 n_1 a_2^\dagger +n_4 a_3 a_1^\dagger+ a_4 a_3^\dagger a_2 a_1^\dagger\)\,,\\
 \hat s_{\tiny \yng(1,1,1)} =\,& a_4^\dagger n_3 n_2 a_1+q \(n_4n_3a_2a_1^\dagger +a_4 a_3^\dagger n_2 n_1 +n_4 a_3 n_1 a_2^\dagger\)\,,\\
 \hat s_{\tiny \yng(1,1,1,1)} =\,& q n_4 n_3 n_2 n_1\,.
}
Note that these are obtained from the above with the replacement $\nu_i \to n_i$. One can similarly work out the case for all other $\hat s_\mu$ but we do not write them here. Note that $\hat s_k=\hat h_k$. As we show in the Appendix, the  operators $\hat h_k$ can also be obtained from a “Lax-like” fermionic operator in the algebraic Bethe ansatz formalism. To check \eqref{ssCq} explicitly, let us order the basis as  
\equ{
\(\emptyset,{\tiny \yng(1)},{\tiny \yng(2)},{\tiny \yng(1,1)},{\tiny \yng(2,1)},{\tiny \yng(2,2)}\)\,.
}
 Then, one can check that explicitly acting with with the operators $\hat x_\nu$ one finds the complete action\footnote{Note the $\hat s_\mu$  move fermions a certain number of times in quantum superposition. However, fermions never over each other and as a consequence there can never appear any minus signs in this expansion, consistent with the fact that the Littlewood-Richardson coefficients are non-negative. }
\eqss{
\hat  s_{\emptyset}=\,&\left(
\begin{array}{cccccc}
 1 & 0 & 0 & 0 & 0 & 0 \\
 0 & 1 & 0 & 0 & 0 & 0 \\
 0 & 0 & 1 & 0 & 0 & 0 \\
 0 & 0 & 0 & 1 & 0 & 0 \\
 0 & 0 & 0 & 0 & 1 & 0 \\
 0 & 0 & 0 & 0 & 0 & 1 \\
\end{array}
\right)\,,\quad 
\hat  s_{\tiny \yng(1)}= \left(
\begin{array}{cccccc}
 0 & 0 & 0 & 0 & q & 0 \\
 1 & 0 & 0 & 0 & 0 & q \\
 0 & 1 & 0 & 0 & 0 & 0 \\
 0 & 1 & 0 & 0 & 0 & 0 \\
 0 & 0 & 1 & 1 & 0 & 0 \\
 0 & 0 & 0 & 0 & 1 & 0 \\
\end{array}
\right)\,, \qquad 
\hat  s_{\tiny \yng(2)}= \left(
\begin{array}{cccccc}
 0 & 0 & 0 & q & 0 & 0 \\
 0 & 0 & 0 & 0 & q & 0 \\
 1 & 0 & 0 & 0 & 0 & 0 \\
 0 & 0 & 0 & 0 & 0 & q \\
 0 & 1 & 0 & 0 & 0 & 0 \\
 0 & 0 & 1 & 0 & 0 & 0 \\
\end{array}
\right)\,, \\
\hat  s_{\tiny \yng(1,1)}=\,& \left(
\begin{array}{cccccc}
 0 & 0 & q & 0 & 0 & 0 \\
 0 & 0 & 0 & 0 & q & 0 \\
 0 & 0 & 0 & 0 & 0 & q \\
 1 & 0 & 0 & 0 & 0 & 0 \\
 0 & 1 & 0 & 0 & 0 & 0 \\
 0 & 0 & 0 & 1 & 0 & 0 \\
\end{array}
\right)\,,\quad 
\hat s_{\tiny \yng(2,1)}=\left(
\begin{array}{cccccc}
 0 & q & 0 & 0 & 0 & 0 \\
 0 & 0 & q & q & 0 & 0 \\
 0 & 0 & 0 & 0 & q & 0 \\
 0 & 0 & 0 & 0 & q & 0 \\
 1 & 0 & 0 & 0 & 0 & q \\
 0 & 1 & 0 & 0 & 0 & 0 \\
\end{array}
\right)\,,\quad 
\hat  s_{\tiny \yng(2,2)}= \left(
\begin{array}{cccccc}
 0 & 0 & 0 & 0 & 0 & q^2 \\
 0 & q & 0 & 0 & 0 & 0 \\
 0 & 0 & 0 & q & 0 & 0 \\
 0 & 0 & q & 0 & 0 & 0 \\
 0 & 0 & 0 & 0 & q & 0 \\
 1 & 0 & 0 & 0 & 0 & 0 \\
\end{array}
\right)\,.
}
These matrix elements precisely match the $q$-deformed Littlewood-Richardson coefficients.\footnote{These can be found for this example in, e.g.,  page 16 of \cite{Gu:2020zpg} as these describe the quantum cohomology ring of the Grassmannian $G(2,4)$.} In the special case $q=0$ the matrix elements exactly reproduce the Littlewood-Richardson coefficients. As we discuss below the case $q=-1$ is special.

\section{Diagonalization by Bethe ansatz}
\label{sec:Diagonalization by Bethe ansatz}

In this section, we show that quantized Schur functions are diagonalized by the Bethe ansatz for the XX spin chain. As a direct consequence we derive a number of formulas involving Schur polynomials evaluates at root of unity, including the celebrated formula by Bertram-Vafa-Intriligator formula \cite{vafa1991topological,INTRILIGATOR_1991,bertram1994schubert,siebert1994quantum} for the ($q$-deformed) Littlewood-Richardson coefficients at $q=(-1)^{k-1}$. We also discuss possible implications for classical and quantum algorithms for approximating these coefficients. 

\subsection{Bethe ansatz and the compound DFT }

Consider the XX spin chain in the magnon sector $k$ (or, in fermionic language, fermion number $k$) with periodic boundary conditions. The Bethe vacua are given by
\equ{\label{Betherootsunity}
\cV_{k,n} = \{(x_1,\ldots, x_k)\; |\; x_i^n=1\,,  x_i\neq x_j\}/S_n\,.
}
Let $\{x_a\}$,  $a=(1,\ldots, {n \choose k})$, denote the set of Bethe roots. Then, each $x_a=e^{ip_a}$ defines a corresponding Bethe state \eqref{Bethestatecoord}. In fact, this is a complete basis for the space of $k$ fermions occupying $n$ fermionic modes.  We denote the  corresponding change of basis matrix from the computational basis to the Bethe basis by $B_k$. Note that an element $x_a\in \cV_{k,n}$ is labeled by an $n$-bit string of Hamming weight $k$, which defines a  partition fitting inside a box of size $k\times (n-k)$. Recall we denoted such set of partitions by $P_{k,n}$. Then, there is an isomorphism
\equ{
P_{k,n}\simeq \cV_{k,n}\,.
}
This matrix change of basis admits a simple description. Consider first the sector of Hamming weight $k=1$, i.e., a single particle on closed chain of $n$ sites and periodic boundary conditions. In this sector the free-fermion Hamiltonian with periodic boundary conditions is diagonalized by the discrete Fourier transform (DFT) matrix:
\equ{\label{Fnmatrix}
 F_n=\frac{1}{\sqrt{n}} 
\begin{pmatrix}
1 & 1 & 1 & \cdots & 1 \\
1 & \omega & \omega^2 & \cdots & \omega^{n-1} \\
1 & \omega^2 & \omega^4 & \cdots & \omega^{2(n-1)} \\
\vdots & \vdots & \vdots & \ddots & \vdots \\
1 & \omega^{n-1} & \omega^{2(n-1)} & \cdots & \omega^{(n-1)^2}
\end{pmatrix} \,,
}
where $\omega=e^{\frac{2\pi i }{n}}$ is the principal $n$th root of unity. Since the Hamiltonian describes non-interacting fermions, in the sector of $k$ fermions one simply goes to to the Fourier basis for each particle, totally antisymmetrized to account for fermionic statistics. Thus, the change of basis is given by the compound of the DFT matrix:
\equ{
B_k= C_k[F_n]
}
where $C_k[M]$ stands for the $k$th compound of the matrix $M$, i.e., the $n \choose k$-dimensional matrix with matrix elements given by all minors of $M$, i.e., $(C[M])_{\mu \nu}=\det M_{\mu \nu}$ where $M_{\mu\nu}$ stands for the submatrix of $M$ obtained by restricting to the rows (and columns) specified by the nonzero entries of the binary strings $\mu$ (and $\nu$) of Hamming weight $k$.  As we show in the Appendix, such matrix elements are given by Schur polynomials evaluated at the $k$-subsets of roots of unity, times a Vandermonde determinant:
\equ{\label{BkVS}
B_k=\frac{1}{n^{k/2}}\sum_{\mu,a} V(x_a)s_\mu(x_a)\ket{a}\bra{\mu}\,.
}
Since the Fourier matrix \eqref{Fnmatrix} is unitary and the compound of a unitary matrix is unitary,  it follows that $B_k$ is unitary:
\equ{
B_kB_k^\dagger= B_k^\dagger B_k=1\,.
}
The matrix change of basis in the full Hilbert space is block diagonal, $B=\bigoplus_k B_k$, and is thus also unitary. An immediate consequence of the unitary of $B_k$ is the following:
\begin{cor} \label{cororthoschur}
The Schur functions satisfy the orthogonality relations
\eqs{
\frac{1}{n^k} \sum_{a\in \cV_{k,n} } \abs{V(x_a)}^2 s_\mu(x_a) s_\nu(x_a^\ast)=\,&\delta_{\mu \nu}\,,\\
\frac{1}{n^k} \sum_{\mu\in P_{k,n} } V(x_b^\ast)V(x_a)  s_\mu(x_a) s_\mu(x_b^\ast)=\,&\delta_{a b}\,.
}
\end{cor}
\noindent These  formulas have appeared in \cite{rietsch2001quantum}. We also note in passing that 
\equ{
V(x_a)S_\mu(x_a)=V(x_\mu)S_a(x_\mu)\,,
}
which follows directly from the fact that since the DFT matrix is symmetric, and the compound of a symmetric matrix is symmetric. Thus, there is a symmetry under exchange of partitions $\mu$ and subsets of roots of unity $a$. 

\begin{claim}
The quantized Schur functions are simultaneously diagonalized in the Bethe basis:
\equ{\label{diagonalquantumschur}
\hat s_\lambda=  B^\dagger D_\lambda B\,,
}
with $D_\lambda$ a diagonal matrix with elements $s_\lambda(x_a)$, with  $a$ running over all possible Bethe roots, 
\equ{
D_\lambda = \sum_{a}s_\lambda(x_a)\ket{a}\bra{a}\,.
} 
\end{claim}
Note this is a way to understand the main result of \cite{fomin1998noncommutative} that the quantized Schur expand the same way as their classical counterparts; in the diagonal basis these statements are equivalent. 

\paragraph{The Bertram-Vafa-Intriligator residue formula.} As a consequence of this formalism a new proof of the well known Bertram-Vafa-Intriligator formula  \cite{vafa1991topological,INTRILIGATOR_1991,bertram1994schubert} follows:
\begin{thm} [Bertram-Vafa-Intriligator] The $q$-deformed Littlewood-Richardson coefficients at $q=(-1)^{k-1}$ are given by 
\equ{
C_{\mu\nu}^{(q)\; \lambda}= \frac{1}{n^k} \sum_{x_a\in V_{k,n}} \abs{V(x_a)}^2\, s_{\mu}(x_a)s_{\nu}(x_a)s_{\lambda}(x_a^\ast)\,. 
}
\end{thm}
\begin{proof}
Follows directly from combining  \eqref{ssCq}, \eqref{diagonalquantumschur} and the orthogonality relations in Corollary~\ref{cororthoschur}. 
\end{proof}

\paragraph{The action of the compound DFT and polynomial multiplication.} To better understand the action of the compound DFT, consider the following. Let $p(\mathbf{ x})$ be a polynomial in $k$ variables $\mathbf{x}=(x_1,\ldots,x_k)$, with an expansion in the Schur basis given by
\equ{p(\mathbf{ x})=\sum_{\mu} \alpha_\mu s_\mu(\mathbf{ x})\,,
}
with the $\alpha_\mu\in \Bbb C$ some coefficients. In quantum mechanical notation one can represent the polynomial in this basis by collecting its coefficients in the Schur basis into the normalized state:
\equ{
 \ket{p}_s=\frac{1}{\norm{\alpha}}\sum_{\mu}\alpha_\mu \ket{\mu}\,,
}
where $\norm{\alpha}=(\sum_\mu \alpha_\mu^2)^{1/2}$.  Let us see the action of the compound DFT on this state. Applying \eqref{BkVS} to both sides and using the orthogonality property in Corollary~\ref{cororthoschur} one gets 
\equ{
\ket{p}_p:=C[F_n] \ket{p}_s=\frac{1}{n^{k/2}} \frac{1}{\norm{\alpha}}\sum_a V(x_a)p(x_a)\ket{a}\,.
} 
This corresponds to a point representation of the polynomial $p(x)$, multiplied by the Vandermonde determinant. Thus, the compound DFT is a map from the description of a polynomial as a list of coefficients in the Schur basis to a point representation, evaluated at subsets of roots of unity. 

\

Note that in the sector of Hamming weight $k=1$ the compound DFT coincides with the action of the classical DFT.  Recall that the DFT is used in the classical setting for the fast multiplication of polynomials. Thus, the compound DFT circuit (see Figure~\ref{fig:CircuitButterfly}) can  be thought of as a “Schur uplift” of the FFT, relevant to the product of Schur polynomials. We comment more on the quantum circuit for the compound DFT below.

\paragraph{Comments on Hamiltonians and quantum circuits.} For each quantized symmetric function above, one can take the real (or imaginary) part to obtain a corresponding hermitian operator, e.g.,
\equ{
\Pi_\lambda:=\hat p_\lambda+\hat p_\lambda^\dagger\,,\qquad \qquad H_\lambda:=\hat h_\lambda+\hat h_\lambda^\dagger\,,\qquad \qquad S_\lambda:=\hat s_\lambda+\hat s_\lambda^\dagger\,.
}
Note the free fermion Hamiltonian is given by $H_{XX}=\Pi_{(1)}$. For a general partition with number of boxes $\abs{\lambda}=d$, however, these are nonlinear polynomials of degree $d$ in the bilinears $a_i^\dagger a_j$ and describe highly interacting fermions. However, these have a very particular structure and since the basis that diagonalizes them is the compound DFT one has a unitary circuit of the form, e.g., 
\equ{
e^{it S_\lambda}= C[F_n]^\dagger e^{it \theta_{\lambda}}C[F_n]\,,
}
where $e^{it \theta_{\lambda}}=\text{diag}\(e^{i (s_\lambda(x_a)+s_\lambda(x_a^\ast)})\)$ is a diagonal unitary,  and similarly for the quantized power sum and complete homogeneous operators. Thus, implementing these unitaries amounts to implementing the diagonal part.

\subsection{A natural generalization of IQP circuits}
\label{sec:A natural generalization of IQP circuits}

\begin{figure}[]
\begin{center}
\includegraphics{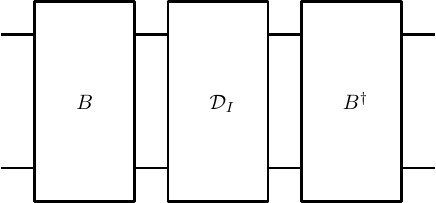}
\caption{A class of quantum circuits consisting of a circuit $B$ implementing the Bethe basis and a circuit $D_I$ that is diagonal in the Z basis. We note the similarly with the class of IQP introduced in \cite{Bremner_2010}, where the first layer in that case is simply a tensor product of Hadamards, $H^{\otimes n}$. Although these are not universal circuits, it is believed that they cannot be exactly simulated classically; see \cite{Bremner_2016} for a discussion of when noise is included. }
\label{fig:Bethecircuits}
\end{center}
\end{figure}

Our discussion above suggests defining a class of quantum circuits relevant to quantum spin chains or, more generally, to any quantum integrable system. Consider the XX quantum spin chain (equivalently, free fermions) and the set of unitary operators $U_I=e^{i(\cO_I+\cO_I^\dagger)}$, with  $\cO_I$ the set of commuting Bethe operators diagonalized in the compound DFT basis $B=C[F_n]$. Then 
\equ{\label{UBDB}
U_I = B^\dagger D_I B\,,
}
with $D_I$ a unitary diagonal matrix. This leads us to consider a  larger class of quantum circuits with this factorized form (see  Figure~\ref{fig:Bethecircuits}) and $D_I$ an {\it arbitrary} diagonal quantum circuit. 
We note that this is analogous to the  “instantaneous quantum polynomial”  (IQP) class of circuits of the form $H^{\otimes n} D_I H^{\otimes n}$, introduced in \cite{Bremner_2010}. Thus, one may refer to the class of quantum circuits defined above ``fermionic IQP'' circuits.\footnote{Recall that IQP circuits can be motivated by the considering quantum circuits of the form {\it classical} Ising spin model at imaginary temperature,  $\text{Tr}\, e^{i\theta H_{\text{Ising}}}$ where $H_{\text{Ising}}$ is the Hamiltonian for the classical Ising model, with arbitrary coupling constants. Note that we have considered here spin chains with homogeneous, and fixed,  coupling constants $\hat J= J_{X,Y,Z}$. However, one can also consider site-dependent coupling constants $\vec J_i$ as well as turn on background magnetic fields, leading to a setting like that of \cite{Bremner_2010,Bremner_2016}.}  

\begin{figure}[]
\begin{center}
\includegraphics{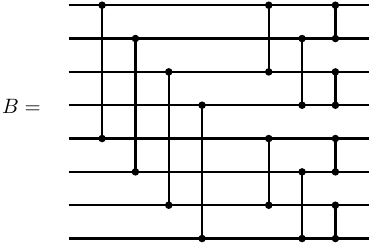}
\caption{The compound DFT circuit relevant to the XX spin chain for $n=8$. This circuit was considered previously in \cite{jain2023quantum} where each 2-qubit gates corresponding to fermionic beam-splitter gates with a fixed angle provided by DFT. The circuit has the same structure as the famous butterfly circuit used for the FFT of \cite{cooley1965algorithm} used for fast multiplication of polynomials, but with fermionic beam splitters for the 2-qubit gates. In the sector of Hamming weight $k=1$ this coincides with the action of the FFT. The circuit can then be thought of as a “Schur uplift” of the FFT, relevant to the product of Schur polynomials. }
\label{fig:CircuitButterfly}
\end{center}
\end{figure}

As we have discussed, for fermionic IQP circuits the outer layer $B=C[F_n]$ is given by the compound DFT. An efficient implementation for this circuit is given by the classical FFT butterfly circuit  of \cite{cooley1965algorithm}, but with the 2-qubit gates replaced by fermionic beam splitters  \cite{jain2023quantum}  (see Figure~\ref{fig:CircuitButterfly}).  Note that although the outer layers in standard IQP circuits can be simulated classically, it is believed that (noiseless) IQP circuits cannot be simulated classically (see \cite{Bremner_2016} for an analysis of noisy cases). Similarly, although the outer layers of fermionic IQP circuits are given by the classically-simulable matchgates, it is possible that fermionic IQP circuits cannot be simulated classically for {\it arbitrary} diagonal circuits $D_I$. It would be interesting to investigate this further.  Note that IQP circuits were  implemented recently on logical qubits in the  beautiful neutral atom demonstrations of \cite{Quera2023}.

More generally, given a quantum integrable system with Bethe basis $B$, one can define the class of unitaries that are diagonalized in this basis. If these admit an efficient implementation, one may refer to these as the class of quantum integrable (QI) circuits.\footnote{To the best of our knowledge, it is not known under which conditions the operators $B$ be implemented efficiently, apart from the case of the XX spin chain discussed above. See \cite{PRXQuantum.2.040329,Van_Dyke_2022,sopena2022algebraic,ruiz2023bethe,raveh2024deterministic} for a discussion of Bethe basis circuits for XXZ spin chains.}

\section{Discussion and open problems}
\label{sec:Discussion and open problems}

We have shown that quantum spin chains naturally “want” to compute a number of nontrivial combinatorial, group theoretic, and geometric quantities. In particular, natural operations on the Hilbert space of  XX quantum spin chain encode the characters of the symmetric group, Kostka coefficients, and $q$-deformed Littlewood-Richardson coefficients.  This is uncovered by constructing appropriate operators on the Hilbert space of the quantum spin chain using the formalism of quantized symmetric functions. The commuting nature of these operators is closely related to the quantum integrability of the system. Indeed, these operators are all diagonalized by the Bethe ansatz  basis for the underlying quantum spin chain.  As a side result this leads to a new proof of the celebrated Bertram-Vafa-Intriligator residue formula for $q$-deformed Littlewood-Richardson coefficients at $q=(-1)^{k-1}$. The formalism of quantized symmetric functions, combined with  the Bethe ansatz, not only sheds new light on various results scattered in the mathematical literature but also suggests possible avenues for developing novel quantum algorithms. As we discussed, in the simplest case of the XX spin chain this formalism naturally leads to the compound DFT transform discussed in \cite{jain2023quantum}. The preparation of more general Bethe states on quantum computers has been considered in \cite{PRXQuantum.2.040329,Van_Dyke_2022,sopena2022algebraic,ruiz2023bethe,raveh2024deterministic}. 

\

As discussed in the Introduction, our motivation is to propose a systematic, bottom-up, approach to identifying computational problems embedded {\it inside} quantum mechanical systems with potential quantum advantage and which are of relevance {\it outside} of quantum mechanics. A systematic approach should start from the simplest (even non-interacting) systems, then identifying  the problems arising there, and work up towards strongly interacting quantum systems where computation is more of a challenge to classical methods. It is rather remarkable in fact that already the simplest quantum mechanical systems reveal nontrivial problems with quantum advantage (boson sampling). It would be interesting to define precise approximation or sampling problems related quantum spin chains and other quantum integrable systems, many of which capture nontrivial mathematical data, develop quantum algorithms for these tasks, and study the nature of any potential speedups. 

Note that here we have considered the simplest quantum spin chain, the Heisenberg XX spin chain. As one moves upwards towards systems with more complex interactions an obvious next step is that of the calculus of the XXZ spin chain. This introduces fermionic interactions already at the level of the Hamiltonian and thus one may expect the quantities to be encoded by Bethe operators to be “more quantum mechanical” than those of the XX spin chain. It is known from the work of Nekrasov- Shatashvili  \cite{nekrasov2009supersymmetric,nekrasov2009quantum}  and Okounkov\cite{GORBOUNOV2017282} that XXZ spin chains capture certain topological invariants known as Gromow-Witten invariants. As far as we know, nothing is known about the computational complexity of these invariants.

Another direction is suggested by the formalism of quantized symmetric functions described here. We have already seen that the regular product of Schur functions leads to Littlewood-Richardson coefficients. In addition to the regular multiplication, however, there are two more natural operations in the ring of symmetric functions,\footnote{In fact, the ring of symmetric functions has the structure of a Hopf algebra. }   Kronecker and plethysm, and one has the relations
\eqs{\label{variousprodsschur1}
s_\mu \ast_q s_\nu =\,& \sum_{\lambda} C_{\mu\nu}^{(q)\;\;\lambda} s_\lambda\,,\\ \label{variousprodsschur2}
s_\mu \ast_K s_\nu =\,& \sum_{\lambda} g_{\mu\nu\lambda} s_\lambda\,,\\ \label{variousprodsschur3}
s_\mu[s_\nu] =\,& \sum_{\lambda} p_{\mu\nu\lambda} s_\lambda\,.
}
The coefficients $g_{\mu\nu\lambda} $ are known as Kronecker coefficients and the $ p_{\mu\nu\lambda}$ are known as plethysm coefficients  \cite{stanley1999enumerative,macdonald1998symmetric}. Although a lot is known about the Littlewood-Richardson coefficients, many important questions remain open about Kronecker and plethysm coefficients; see \cite{stanley2006some} for some important problems in the field and \cite{colmenarejo2022mystery} for a discussion of the mystery of plethysm coefficients.  We have shown how that the ($q$-deformed) Littlewood-Richardson  are captured by quantized Schur functions acting on the XX spin chain.  It would be interesting to investigate whether the formalism of quantized symmetric functions can also shed light on Kronecker and plethysm coefficients and what is the quantum mechanics of these coefficients. The (quantum) complexity of  Kronecker coefficients was recently considered in \cite{bravyi2023quantum}. 

\paragraph{Acknowledgements:} We thank Adam Bouland, Sergei Bravyi, Max Gordon, Vojt\v{e}ch Havl\'ic\v{e}k, Iordanis Kerenidis, Mart\'in Larocca, German Sierra, Alejandro Sopena and Guanyu Zhu for discussions and especially Brian Willett for collaboration at various stages.

\bibliography{references.bib} 

\providecommand{\href}[2]{#2}\begingroup\raggedright\begin{thebibliography}{10}

\bibitem{365700}
P.~Shor, \href{http://dx.doi.org/10.1109/SFCS.1994.365700}{``Algorithms for
  quantum computation: discrete logarithms and factoring,''} in {\em
  Proceedings 35th Annual Symposium on Foundations of Computer Science},
  pp.~124--134.
\newblock 1994.

\bibitem{childs2010quantum}
A.~M. Childs and W.~Van~Dam, ``Quantum algorithms for algebraic problems,''
  {\em Reviews of Modern Physics} {\bf 82} (2010) no.~1, 1.

\bibitem{montanaro2016quantum}
A.~Montanaro, ``Quantum algorithms: an overview,'' {\em npj Quantum
  Information} {\bf 2} (2016) no.~1, 1--8.

\bibitem{aaronson2010computational}
S.~Aaronson and A.~Arkhipov, ``The computational complexity of linear optics,''
  2010.

\bibitem{stanley1999enumerative}
R.~Stanley and S.~Fomin, {\em Enumerative Combinatorics: Volume 2}.
\newblock Cambridge Studies in Advanced Mathematics. Cambridge University
  Press, 1999.
\newblock \url{https://books.google.com/books?id=cWEhAwAAQBAJ}.

\bibitem{macdonald1998symmetric}
I.~G. Macdonald, {\em Symmetric functions and Hall polynomials}.
\newblock Oxford university press, 1998.

\bibitem{egge2019introduction}
E.~S. Egge, {\em An introduction to symmetric functions and their
  combinatorics}, vol.~91.
\newblock American Mathematical Soc., 2019.

\bibitem{gelfand1994noncommutative}
I.~Gelfand, D.~Krob, A.~Lascoux, B.~Leclerc, V.~S. Retakh, and J.~Y. Thibon,
  ``Noncommutative symmetric functions,'' 1994.

\bibitem{fomin1998noncommutative}
S.~Fomin and C.~Greene, ``Noncommutative schur functions and their
  applications,'' {\em Discrete Mathematics} {\bf 193} (1998) no.~1-3,
  179--200.

\bibitem{bravyi2023quantum}
S.~Bravyi, A.~Chowdhury, D.~Gosset, V.~Havlicek, and G.~Zhu, ``Quantum
  complexity of the kronecker coefficients,'' {\em arXiv preprint
  arXiv:2302.11454} (2023)  .

\bibitem{jordan2008fast}
S.~P. Jordan, ``Fast quantum algorithms for approximating some irreducible
  representations of groups,'' {\em arXiv preprint arXiv:0811.0562} (2008)  .

\bibitem{Childs_2018}
A.~M. Childs, D.~Maslov, Y.~Nam, N.~J. Ross, and Y.~Su,
  \href{http://dx.doi.org/10.1073/pnas.1801723115}{``Toward the first quantum
  simulation with quantum speedup,''{\em Proceedings of the National Academy of
  Sciences} {\bf 115} (Sept., 2018)  9456–9461}.
  \url{http://dx.doi.org/10.1073/pnas.1801723115}.

\bibitem{nekrasov2009supersymmetric}
N.~A. Nekrasov and S.~L. Shatashvili, ``Supersymmetric vacua and bethe
  ansatz,'' {\em arXiv preprint arXiv:0901.4744} (2009)  .

\bibitem{nekrasov2009quantum}
N.~Nekrasov and S.~Shatashvili, ``Quantum integrability and supersymmetric
  vacua,'' {\em Progress of Theoretical Physics Supplement} {\bf 177} (2009)
  105--119.

\bibitem{maulik2012quantum}
D.~Maulik and A.~Okounkov, ``Quantum groups and quantum cohomology,'' {\em
  arXiv preprint arXiv:1211.1287} (2012)  .

\bibitem{GORBOUNOV2017282}
V.~Gorbounov and C.~Korff, ``Quantum integrability and generalised quantum
  schubert calculus,''
  \href{http://dx.doi.org/https://doi.org/10.1016/j.aim.2017.03.030}{{\em
  Advances in Mathematics} {\bf 313} (2017)  282--356}.
  \url{https://www.sciencedirect.com/science/article/pii/S0001870817300932}.

\bibitem{zinnjustin2009littlewoodrichardson}
P.~Zinn-Justin, ``Littlewood--richardson coefficients and integrable tilings,''
  2009.

\bibitem{zinn2008integrability}
P.~Zinn-Justin, ``Integrability and combinatorics: selected topics,'' {\em Les
  Houches lecture notes, http://www. lpthe. jussieu. fr/\~{}
  pzinn/semi/intcomb. pdf} {\bf 30} (2008)  .

\bibitem{baxter2016exactly}
R.~J. Baxter, {\em Exactly solved models in statistical mechanics}.
\newblock Elsevier, 2016.

\bibitem{eckle2019models}
H.-P. Eckle, {\em Models of Quantum Matter: A First Course on Integrability and
  the Bethe Ansatz}.
\newblock Oxford University Press, 2019.

\bibitem{sagan2013symmetric}
B.~E. Sagan, {\em The symmetric group: representations, combinatorial
  algorithms, and symmetric functions}, vol.~203.
\newblock Springer Science \& Business Media, 2013.

\bibitem{Gu:2020zpg}
W.~Gu, L.~Mihalcea, E.~Sharpe, and H.~Zou, ``{Quantum K theory of symplectic
  Grassmannians},''
  \href{http://dx.doi.org/10.1016/j.geomphys.2022.104548}{{\em J. Geom. Phys.}
  {\bf 177} (2022)  104548}, \href{http://arxiv.org/abs/2008.04909}{{\tt
  arXiv:2008.04909 [hep-th]}}.

\bibitem{vafa1991topological}
C.~Vafa, ``Topological mirrors and quantum rings,'' 1991.

\bibitem{INTRILIGATOR_1991}
K.~INTRILIGATOR, \href{http://dx.doi.org/10.1142/s0217732391004097}{``Fusion
  residues,''{\em Modern Physics Letters A} {\bf 06} (Dec., 1991)
  3543–3556}. \url{http://dx.doi.org/10.1142/S0217732391004097}.

\bibitem{bertram1994schubert}
A.~Bertram, ``Towards a schubert calculus for maps from a riemann surface to a
  grassmannian,'' 1994.

\bibitem{siebert1994quantum}
B.~Siebert and G.~Tian, ``On quantum cohomology rings of fano manifolds and a
  formula of vafa and intriligator,'' 1994.

\bibitem{rietsch2001quantum}
K.~Rietsch, ``Quantum cohomology of grassmannians and total positivity,'' 2001.

\bibitem{Bremner_2010}
M.~J. Bremner, R.~Jozsa, and D.~J. Shepherd,
  \href{http://dx.doi.org/10.1098/rspa.2010.0301}{``Classical simulation of
  commuting quantum computations implies collapse of the polynomial
  hierarchy,''{\em Proceedings of the Royal Society A: Mathematical, Physical
  and Engineering Sciences} {\bf 467} (Aug., 2010)  459–472}.
  \url{http://dx.doi.org/10.1098/rspa.2010.0301}.

\bibitem{Bremner_2016}
M.~J. Bremner, A.~Montanaro, and D.~J. Shepherd,
  \href{http://dx.doi.org/10.1103/physrevlett.117.080501}{``Average-case
  complexity versus approximate simulation of commuting quantum
  computations,''{\em Physical Review Letters} {\bf 117} (Aug., 2016)  }.
  \url{http://dx.doi.org/10.1103/PhysRevLett.117.080501}.

\bibitem{cooley1965algorithm}
J.~W. Cooley and J.~W. Tukey, ``An algorithm for the machine calculation of
  complex fourier series,'' {\em Mathematics of computation} {\bf 19} (1965)
  no.~90, 297--301.

\bibitem{jain2023quantum}
N.~Jain, J.~Landman, N.~Mathur, and I.~Kerenidis, ``Quantum fourier networks
  for solving parametric pdes,'' 2023.

\bibitem{Quera2023}
D.~Bluvstein, S.~J. Evered, A.~A. Geim, S.~H. Li, H.~Zhou, T.~Manovitz,
  S.~Ebadi, M.~Cain, M.~Kalinowski, D.~Hangleiter, J.~P. Bonilla~Ataides,
  N.~Maskara, I.~Cong, X.~Gao, P.~Sales~Rodriguez, T.~Karolyshyn, G.~Semeghini,
  M.~J. Gullans, M.~Greiner, V.~Vuleti{\'c}, and M.~D. Lukin, ``Logical quantum
  processor based on reconfigurable atom arrays,''
  \href{http://dx.doi.org/10.1038/s41586-023-06927-3}{{\em Nature} {\bf 626}
  (2024) no.~7997, 58--65}. \url{https://doi.org/10.1038/s41586-023-06927-3}.

\bibitem{PRXQuantum.2.040329}
J.~S. Van~Dyke, G.~S. Barron, N.~J. Mayhall, E.~Barnes, and S.~E. Economou,
  \href{http://dx.doi.org/10.1103/PRXQuantum.2.040329}{``Preparing bethe ansatz
  eigenstates on a quantum computer,''{\em PRX Quantum} {\bf 2} (Nov, 2021)
  040329}. \url{https://link.aps.org/doi/10.1103/PRXQuantum.2.040329}.

\bibitem{Van_Dyke_2022}
J.~S. Van~Dyke, E.~Barnes, S.~E. Economou, and R.~I. Nepomechie,
  \href{http://dx.doi.org/10.1088/1751-8121/ac4640}{``Preparing exact
  eigenstates of the open xxz chain on a quantum computer,''{\em Journal of
  Physics A: Mathematical and Theoretical} {\bf 55} (Jan., 2022)  055301}.
  \url{http://dx.doi.org/10.1088/1751-8121/ac4640}.

\bibitem{sopena2022algebraic}
A.~Sopena, M.~H. Gordon, D.~Garc{\'\i}a-Mart{\'\i}n, G.~Sierra, and
  E.~L{\'o}pez, ``Algebraic bethe circuits,'' {\em Quantum} {\bf 6} (2022)
  796.

\bibitem{ruiz2023bethe}
R.~Ruiz, A.~Sopena, M.~H. Gordon, G.~Sierra, and E.~López, ``The bethe ansatz
  as a quantum circuit,'' 2023.

\bibitem{raveh2024deterministic}
D.~Raveh and R.~I. Nepomechie, ``Deterministic bethe state preparation,'' 2024.

\bibitem{stanley2006some}
R.~P. Stanley, ``Some combinatorial aspects of the schubert calculus,'' in {\em
  Combinatoire et Repr{\'e}sentation du Groupe Sym{\'e}trique: Actes de la
  Table Ronde du CNRS tenue {\`a} l'Universit{\'e} Louis-Pasteur de Strasbourg,
  26 au 30 avril 1976}, pp.~217--251.
\newblock Springer, 2006.

\bibitem{colmenarejo2022mystery}
L.~Colmenarejo, R.~Orellana, F.~Saliola, A.~Schilling, and M.~Zabrocki, ``The
  mystery of plethysm coefficients,'' 2022.

\bibitem{olmedilla1987yang}
E.~Olmedilla, M.~Wadati, and Y.~Akutsu, ``Yang-baxter relations for spin models
  and fermion models,'' {\em Journal of the Physical Society of Japan} {\bf 56}
  (1987) no.~7, 2298--2308.

\bibitem{umeno1998fermionic}
Y.~Umeno, M.~Shiroishi, and M.~Wadati, ``Fermionic r-operator for the fermion
  chain model,'' {\em Journal of the Physical Society of Japan} {\bf 67} (1998)
  no.~6, 1930--1935.

\bibitem{Alexandersson2020}
P.~Alexandersson, ``The symmetric functions catalog.'' Online.
\newblock \url{https://www.symmetricfunctions.com}.

\bibitem{eisenbud20163264}
D.~Eisenbud and J.~Harris, {\em 3264 and all that: A second course in algebraic
  geometry}.
\newblock Cambridge University Press, 2016.

\bibitem{panova2023computational}
G.~Panova, ``Computational complexity in algebraic combinatorics,'' 2023.

\bibitem{panova2023complexity}
G.~Panova, ``Complexity and asymptotics of structure constants,'' 2023.

\bibitem{stanley2000positivity}
R.~P. Stanley, ``Positivity problems and conjectures in algebraic
  combinatorics,'' {\em Mathematics: Frontiers and Perspectives: Frontiers and
  Perspectives} (2000)  295.

\bibitem{burgisser2013deciding}
P.~Bürgisser and C.~Ikenmeyer, ``Deciding positivity of littlewood-richardson
  coefficients,'' 2013.

\bibitem{narayanan2006complexity}
H.~Narayanan, ``On the complexity of computing kostka numbers and
  littlewood-richardson coefficients,'' {\em Journal of Algebraic
  Combinatorics} {\bf 24} (2006)  347--354.

\bibitem{Ikenmeyer_2017}
C.~Ikenmeyer, K.~D. Mulmuley, and M.~Walter,
  \href{http://dx.doi.org/10.1007/s00037-017-0158-y}{``On vanishing of
  kronecker coefficients,''{\em computational complexity} {\bf 26} (July, 2017)
   949–992}. \url{http://dx.doi.org/10.1007/s00037-017-0158-y}.

\bibitem{burgisser2008complexity}
P.~B{\"u}rgisser and C.~Ikenmeyer, ``The complexity of computing kronecker
  coefficients,'' {\em Discrete Mathematics \& Theoretical Computer Science}
  (2008) no.~Proceedings, .

\bibitem{fischer2020computational}
N.~Fischer and C.~Ikenmeyer, ``The computational complexity of plethysm
  coefficients,'' 2020.

\bibitem{ikenmeyer2022positivity}
C.~Ikenmeyer, I.~Pak, and G.~Panova, ``Positivity of the symmetric group
  characters is as hard as the polynomial time hierarchy,'' 2022.

\end{thebibliography}\endgroup
\bibliographystyle{utphys}

\appendix

\section{The Bethe ansatz and quantum integrability}
\label{app:The Bethe ansatz and quantum integrability}

There are various formulations of the Bethe ansatz. In the main text we reviewed the basic elements of the coordinate Bethe ansatz. It turns out that the Bethe states \eqref{Bethestatecoord} not only diagonalize the Hamiltonian but a large number of other physical observables. A large number of mutually commuting observables can be considered as the defining property of  “quantum integrable systems.” A formalism that makes this more transparent and that leads to a general formalism for the study of quantum integrable systems is the algebraic Bethe ansatz formalism.

\subsection{The algebraic Bethe ansatz}

The basic tool of the algebraic Bethe Ansatz approach is the Lax operator. In this formalism one introduces an auxiliary space $V_a$ which in this case is $\Bbb C^2$. The Lax operators $L_i$ are linear operators, acting on the Hilbert space $V_i=\Bbb C^2_i$ at site $i$ and the auxiliary space, i.e., $L_{i,a}:  V_i\otimes V_a\to  V_i\otimes V_a $. The Lax operator for the XXZ spin chain is 
\equ{
L_{i,a} = \begin{pmatrix} u+i S_{i}^z & i S_i^-\\ i S_i^+& u-i S_i^z \end{pmatrix}\,.
} 
This can be written as 
\equ{\label{defLax}
L_{i,a}(u)= \(u-\frac{i}{2}\)\Bbb I + i P_{i,a}
}
where $P_{i,a}$ is the permutation operator, acting as $P(x\otimes y)=y\otimes x$. Another important operator is the quantum R-matrix, $R_{ab}:  V_a\otimes V_a\to  V_a\otimes V_a$, given by 
\equ{\label{defR}
R_{ab}(\lambda)= \lambda\,  \Bbb I + i P_{ab}\,.
}
Note this the form of this operator is basically identical to the Lax operator, but these two operators act on different spaces.  The crucial properties of these operators is that they satisfy the commutation relation:
\equ{\label{RLL}
R_{ab}(u_1-u_2)L_{i,a}(u_1)L_{i,b}(u_2)=L_{i,b}(u_2)L_{i,a}(u_1)R_{ab}(u_1-u_2)
}
as can be easily checked using the definitions \eqref{defLax} and \eqref{defR}. Then, one defines the monodromy matrix
\equ{
M_a(u) := L_{n,a}(u) L_{n-1,a}(u)\cdots L_{1,a}(u)\,.
}
Note that each $L_{i,a}(u)$ can be written as a $2\times2$ matrix corresponding to the auxiliary space, with each entry acting on the physical space $V_i$. Then, the monodromy matrix has the form
\equ{
M(u) = \begin{pmatrix} A(u) & B(u)\\ C(u) & D(u) \end{pmatrix}\,,
}
where each entry are operators in the physical space $\cH=\bigotimes_i V_i$.  The transfer matrix is defined as the $q$-trace of the monodromy matrix: 
\equ{
T_q(u) := q A(u) + D(u)\,.
}
Now, the claim is using the commutation relations \eqref{RLL} and cyclicity of the trace, the transfer matrices commute:
\equ{
[T_q(u_1),T_q(u_2)]=0\,.
}
Note that each $T_q(u)$ can be expanded in powers of $u$,
\equ{
T_q(u) = \sum_{i=1}^n u^i T_i\,.
}
This implies that all the $T_i$ are a collection of mutually  commuting operators:
\equ{
[T_i,T_j]=0\,.
}
Thus, the operators can be all simultaneously diagonalized.\footnote{In fact, these operators are an abelian subset of a larger set of operators $T_{ij}$ which satisfy an algebra known as the Yangian, and it this extended symmetry which is crucial property of quantum integrable systems.   } In fact, they are diagonalized by the so-called Bethe states, which are obtained by acting with the operators $C$ in the monodromy matrix. Defining
\equ{\label{BetheStates}
\ket{u_1,\ldots,u_k} = B(u_k)\cdots B(u_1) \ket{0^n}\,,
}
one can show that 
\equ{
T_i \ket{u_1,\ldots,u_k}   = \lambda_{i} \ket{u_1,\ldots,u_k} 
}
provided the $u_i$ satisfy the Bethe equations \eqref{betheeqs}. 

\subsection{Quantized symmetric functions}

Since the quantized symmetric functions introduced above  all commute with one another, it is natural to wonder whether these have a natural description in the algebraic Bethe formalism. Let us consider the functions $\hat h_\lambda$. Note that these are expressed in terms of fermionic operators rather than spin operators. That is, to compare those operators to the standard operators of the XX spin chain one has to reformulate the formalism above in terms of fermions. Here, we just make the following observation. Define the following fermionic operators at each site:
\equ{\label{LhatL}
 L^F_i(u) := \begin{pmatrix} u\, \nu_i & a_i\\ u\, a_i^\dagger & 1 \end{pmatrix}\,,
 }
and the product over the chain:
\equ{
M^F_a(u) :=   L^F_{n,a}(u)   L^F_{n-1,a}(u)\cdots   L^F_{1,a}(u)= \begin{pmatrix}  A_F(u) &  B_F(u)\\  C_F(u) &  D_F(u) \end{pmatrix}\,.
}
Taking the $q$-trace of this operator we find
\equ{
 T^F_q(u) := q  A_F(u) +  D_G(u) = \sum_{k}u^k  \hat s_k\,,
}
where the $\hat s_k$ are the $q$-deformed quantized Schur functions in \eqref{defShat} in the fermionic representation \eqref{defxi}. We note that the operators \eqref{LhatL} are similar, but not identical, to the fermionic operators discussed in \cite{olmedilla1987yang,umeno1998fermionic}. It would be interesting to understand the relation between these in more detail.

\section{Symmetric functions}
\label{App:Symmetric functions}

 The theory of symmetric functions has a number of applications in  combinatorics, group theory, Lie algebras, and algebraic geometry. Here we review some basic notions and definitions. Standard references are the books by Stanley \cite{stanley1999enumerative}  and  Macdonald \cite{macdonald1998symmetric}. A more introductory book is by Egge \cite{egge2019introduction}.\footnote{See  \cite{Alexandersson2020} for a catalog of symmetric functions and many of its properties.} 

\

Let $\mathbf{x}=(x_1,x_2,\ldots)$ be an infinite set of variables. A {\it homogeneous symmetric function of degree $d$} is a formal power series
\equ{
f(\mathbf{x})=\sum_\alpha c_\alpha x^\alpha\,,
}
where $\alpha$ ranges over all compositions $\alpha=(\alpha_1, \alpha_2, \ldots)$ of $d$,  $x^\alpha=x_1^{\alpha_1}x_2^{\alpha_2}\cdots$, and the $c_\alpha$ are coefficients in a field $R$. The function is homogeneous since $\sum_i \alpha_i=d$ and is symmetric  if
\equ{
f(x_1,x_2,\ldots)=f(x_{w(1)},x_{w(2)},\ldots )\,,
}
for every permutations $w$ of the integers. Let $\Lambda^d_R$ denote the set of homogeneous functions of degree $d$ over a field $R$. Here we will take $R=\Bbb C$ and simply write $\Lambda^d$. 

The set of all such polynomials with arbitrary degree is denoted $\Lambda = \cup_d \Lambda^d$. Note that $f(\mathbf{x})$ is a formal power series, generically containing an infinite number of terms. Note that $\Lambda^d$ has a vector space structure, since the addition of symmetric functions and their multiplication by a scalar are in $\Lambda^d$. The multiplication of two functions of degree $d$ and $d’$ leads to a symmetric function of degree $dd’$.  Thus,  
\equ{
\Lambda = \cup_d \Lambda^d = \text{Sym}\(\Bbb C[x_1,x_2,\ldots]\)
}
has a ring structure as is known as the ring of symmetric functions. In fact, $\Lambda$ has an even richer structure known as as Hopf algebra which we review below.

A simple way to obtain symmetric functions of degree $d$ is as follows. Consider a partition $\lambda = (\lambda_1,\lambda_2, \ldots, \lambda_l)  \vdash d$, where $l\leq d$. Then,  the symmetric polymomial $m_{\lambda}(\mathbf{x})\in \Lambda^d$ is defined by 
\equ{
m_\lambda(\mathbf{x}) :=\sum_{\alpha}x^\alpha\,,
}
where the sum is over all {\it distinct} permutations $\alpha=(\alpha_1, \alpha_2, \ldots)$ of the entries of $\lambda = (\lambda_1,\lambda_2, \ldots)$. For instance, for $d=3$ the partitions $\lambda=(3)$, $\lambda=(2,1)$, and $\lambda=(1,1,1)$ define 
\eqs{
m_3(\mathbf{x}) =\,& x_1^3+x_2^3+x_3^3+x_4^3+\cdots \\
m_{21}(\mathbf{x}) =\,& x_1^2 x_2+x_1^2 x_3+x_1^2 x_4+x_2^2x_3+\cdots \,, \\
m_{111}(\mathbf{x}) =\,& x_1 x_2 x_3+x_1 x_2x_4+x_2x_3x_4+\cdots \,,.
}
It turns out that the set $\{m_\lambda\;|\;\lambda \vdash d\}$ forms a basis for $\Lambda^d$, i.e., any function $f(\mathbf{x})\in \Lambda_R^d$ can be written (uniquely) as
\equ{
f(\mathbf{x})=\sum_{\lambda\vdash d} c_\lambda m_\lambda(\mathbf{x})\,,
}
for some coefficients $c_\lambda\in R$.  The basis $\{m_\lambda\}$ is known as the monomial base for symmetric functions. Although this is perhaps the most natural basis, one can define other basis which have a number of interesting properties. Four standard bases are the elementary basis, $\{e_\lambda\}$, the complete basis, $\{h_\lambda\}$, the power sum basis, $\{p_\lambda\}$, and the Schur basis, $\{s_\lambda\}$.  A basic problem in the theory of symmetric functions is, given a symmetric function $f(\mathbf{x})$ in one basis, to find the coefficients in another basis. It turns out that these changes of basis encode solutions to combinatoric and group theoretic problems, as we review below.

\subsection{The $(e,h,p,s)$ bases}

We begin by describing the various bases for the ring of symmetric functions and their properties. 

\paragraph{The elementary, complete, and power sum bases.}
 For any $k\geq 1$ one defines (see  \cite[Section 7.7]{stanley1999enumerative}), 
\eqs{
e_r :=\,&\sum _{1\leq j_{1}<j_{2}<\cdots <j_{r}}x_{j_{1}}\dotsm x_{j_{r}}\,,\\
h_r:=\,&\sum _{1\leq i_{1}\leq i_{2}\leq \cdots \leq i_{r}}x_{i_{1}}x_{i_{2}}\cdots x_{i_{r}}\,,\\
p_r:=\,&\sum_{1\leq i} x_i^r\,.
}
These should be thought of as formal power series with infinitely many terms. It is convenient to define $e_0=h_0=1$. Then, consider any $\lambda=(\lambda_1,\lambda_2,\cdots,\lambda_l)$ a partition of $d$, and define
\eqs{
e_{\lambda}:=\,&e_{\lambda_1} e_{\lambda_2}\cdots  e_{\lambda_l} \,,\\
h_{\lambda}:=\,&h_{\lambda_1} h_{\lambda_2}\cdots  h_{\lambda_l} \,,\\
 p_\lambda :=\,& p_{\lambda_1}p_{\lambda_2}\cdots  p_{\lambda_l}\,.
}
It turns out that these are also all bases for symmetric functions of degree $d$, 
\equ{
\Lambda^d=\text{span}\{e_\lambda\}=\text{span}\{h_\lambda\}=\text{span}\{p_\lambda\}\,, \qquad \lambda \vdash d\,.
}
Note also that each $e_{\lambda},h_{\lambda},p_{\lambda}$ is given in terms of products of the fundamental polynomials $e_r$, $h_r$, $p_r$. Thus, the ring of symmetric functions can be thought of as
\equ{
\Lambda \simeq \Bbb C[e_1,e_2,\ldots]\simeq \Bbb C[h_1,h_2,\ldots]\simeq \Bbb C[p_1,p_2,\ldots]
}
This is the fundamental theorem of symmetric functions. The fundamental basis has the additional property that
\equ{
\Lambda_\Bbb Z=\Bbb Z[e_1,e_2,\ldots]
}
That is, any symmetric function with integer coefficients in its monomial expansion can be written as a polynomial in the $e_i$ with integer coefficients. This is not the case, for example, for the $p$-basis. 

\paragraph{Schur basis.}

The fundamental combinatorial object associated with Schur functions are semistandard Young tableaux (SSYT).\footnote{Although their definition is less transparent than those of the bases above, their importance arises from their connections with representation theory and algebraic geometry. A discussion of the connection to the representation theory of the symmetric group $S_n$ and the general linear group $\text{GL}(n, \mathbb{C})$ is given in Section 7.18 and Appendix 2 of \cite{stanley1999enumerative}, respectively. The connection to intersection theory on the Grassmannian   is covered  in many places, for instance \cite{eisenbud20163264}} Given a partition $\lambda$, a semistandard Young tableau of shape $\lambda$ is an array $T=(T_{ij})$ of positive integers of shape $\lambda$ that is weakly increasing in every row and strictly increasing in every column. Then, for a partition $\lambda = (\lambda_1, \lambda_2, \ldots, \lambda_n)$, the Schur polynomial is a sum of monomials,
\equ{
 s_\lambda(\mathbf{x})=\sum_{T\in \text{SSYT}(\lambda)} x^T  \,,
}
where the summation is over all SSYTs  $T$ of shape $\lambda$ and the exponents each $t_i$ counts the occurrences of the number $i$ in $T$. In other words, Schur polynomials are the generating functions of SSYTs. As an example, consider the partition $\lambda=(2,1)$, 
\equ{
s_{(2,1)}(\mathbf{x})= x_1^2x_2 + x_2^2x_1+x_1^2x_3+x_1^2x_2+\cdots
}
The collection $\{s_\lambda\;\; \lambda\vdash d\}$ also serves as a basis for $\Lambda^d$.

One can write the sum above as a sum over partitions (rather than SSYT) as
\equ{
s_\lambda(\mathbf{x}) =\sum_{\mu \vdash n}K_{\lambda,\mu} m_\mu(\mathbf{x})
}
where $K_{\lambda\mu}$ is the number of SSYT of shape $\lambda$ and type $\mu$ known as Kostka coefficients. This is an example of a transition matrix between between two bases; we’ll see more below. 

\

Note that 
\equ{
s_k = h_k\,,\qquad  s_{1^k}=e_k\,,\qquad m_{i^k} = e_k \,,\qquad m_k=p_k\,.
}

\subsection{The Hall inner product}
\label{app:The Hall inner product}

One can define an inner product function, $\langle \cdot, \cdot \rangle: \Lambda \times \Lambda \to \Bbb Q $. The inner product is defined by requiring that $\{m_\lambda\}$ and $\{h_\mu\}$ are dual bases, i.e.,
\equ{
\langle m_\lambda, h_\mu\rangle  = \delta_{\lambda \mu}\,,
}
where $ \delta_{\lambda \mu}=1$ if $\lambda =\mu$ and $ \delta_{\lambda \mu}=0$ otherwise. It is straightforward to show that the scalar product is symmetric, i.e., $\langle f,g\rangle = \langle g,f\rangle$ for any $f,g$. It is easy to see that the Schur basis is self dual with respect to the Hall inner product, i.e., 
\equ{\label{orthoschur}
\langle s_\lambda, s_\mu\rangle  = \delta_{\lambda \mu}\,,
}
which follows directly from the transition matrices above and the linearity of the Hall inner product.\footnote{As discussed below, Schur functions can be thought of as the characters of polynomial representations of $GL(n,\Bbb C)$. Then, this orthogonality property is related to the orthogonality of characters.} The inner products between the Schur basis and the other basis are computed similarly, giving 
\equ{
\langle s_\nu,e_\mu\rangle=K_{\mu,\nu^T}\,,\quad \langle s_\nu,h_\mu\rangle=K_{\mu,\nu}\,,\quad \langle s_\nu,m_\mu\rangle=K^{-1}_{\mu,\nu}\,,\quad  \langle s_\nu,p_\mu\rangle=\chi_{\nu}(\mu)\,.
}
A useful set of identities are known as the Cauchy identities \cite{stanley1999enumerative}:
\eqss{
\sum_{\lambda} s_\lambda (x)s_\lambda(y)= \sum_{\lambda} h_\lambda (x)m_\lambda(y)=\sum_{\lambda} z_\lambda^{-1}p_\lambda (x)p_\lambda(y)=\prod_{i,j}\frac{1}{1-x_i y_j}\,,\\
\sum_{\lambda} s_\lambda (x)s_{\lambda^T}(y)=\sum_{\lambda} m_\lambda (x)e_\lambda(y)=\sum_{\lambda} \epsilon_\lambda z_\lambda^{-1}p_\lambda (x)p_\lambda(y)=\prod_{i,j}(1+x_i y_j)
}

%
%

\subsection{More on the Schur basis}

Here we review other expressions for the Schur functions in terms of determinants and in terms of non-intersecting paths on the $\Bbb Z\times \Bbb Z$  lattice.

\paragraph{Weyl and Jacobi-Trudy formulas.}

It is possible to limit the infinite set of variables $\mathbf{x}$ to a finite set $(x_1,\ldots, x_n)$, by formally setting all variables $x_{i}=0$ for all $i>n$. Then,  one writes
\equ{
s_\lambda (x_1,\ldots, x_{n})=s_\lambda (x_1,\ldots, x_{n},0,\ldots)\,,
}
where on the RHS an infinite number of variables have been set to zero. If one is interested in properties of partitions $\lambda$ of a finite size , it is sufficient to work in a finite number of variables, as long as  $n$ is large enough.

Note that Schur functions in a finite number of variables have a “stability” property, namely for any $k\leq n$, 
\equ{
s_\lambda (x_1,\cdots, x_{k},0\cdots,0) =s_\lambda (x_1,\cdots, x_{k}) \,.
}
In what follows we work with a finite number of variables $n$. 

The original definition (sometimes called the classical definition) of Schur functions is via Jacobi's bialternant formula:
\equ{\label{schurJacobi}
s_\lambda(x_1,x_2,\ldots,x_n)=\frac{\det\(x_i^{\lambda_j+n-j}\)_{i\leq n,\, j\leq n}}{\det\(x_i^{n-j}\)_{i\leq i,\, j\leq n }}\,.
}
Note that both numerator and denominator are antisymmetric under exchange of any two variables $x_i$ and $x_j$ and thus the ratio is indeed a symmetric function.\footnote{The denominator can also be written as the Vandermonde determinant, $\det\(x_i^{k-j}\)_{i\leq i,\, j\leq k }=\prod_{1\leq i <j \leq k} (x_i-x_j)$.} Also note that the denominator equals the numerator for $\lambda=0$, implying $s_0=1$ for the empty partition. This is a special case of the Weyl formula (see section on group theory for more details.)
Another expression is in terms of the elementary or complete symmetric functions via the Jacobi-Trudy or Giambelli formula  \cite[Theorem 7.16.1]{stanley1999enumerative}:
\equ{
s_{\lambda/\mu}(x):=\det \(  h_{\lambda_i-\mu_j+j-i} (x) \)=\det \(  e_{\lambda^T_i-\mu^T_j+j-i}(x) \)\,,
}
where $h_0=1$ and $h_i=0$ for $i<0$ and similarly for the $e_i$. Recall the combinatorial definition for Schur fuctions also applies to skew shapes and one can see that the above reproduces the same functions. 

\paragraph{As compound matrices.}

Consider the $n\times n$ Vandermonde matrix
\equ{
X^{(n)} = \begin{pmatrix} 1 & 1 & \cdots & 1 \\ x_0 & x_1 &\cdots &x_{n-1} \\ x_0^2 &x_1^2& \cdots & x_{n-1}^2\\ \vdots &\vdots &\ddots &\vdots \\x_0^{n-1} & x_1^{n-1} &\cdots &x_{n-1}^{n-1} \end{pmatrix}\,, 
}
with determinant
\equ{
V_n:=\det(X^{(n)})=\prod_{0\leq i<j\leq n-1}({x_j-x_i})\,.
}
Now, let us consider Schur functions in $k\leq n$ variables. The claim is that $s_{\lambda}(x_1,\ldots,x_k)$ are obtained from the matrix $X^{(n)}$, as follows. Consider the set 
\equ{
P_{k,n}=\left\{\lambda\;\;\Big|\;\; l(\lambda)\leq k\,,\; w(\lambda)\leq n-k\right\}.
}
This describes the set of partitions $\lambda$ fitting inside a box of size $k\times (n-k)$. Note that for fixed $k,n$ the maximum length and width is fixed, but not the number of boxes. Thus, the set contains all partitions of maximum degree $d\leq k\times (n-k)$. Note that there are $n\choose k$ such partitions. In fact, they can all be represented by the staircase representation:
\equ{
P_{k,n}= \left\{s\in\{0,1\}^n\;\;\Big|\;\; \abs{s}=k\right\}
}
where $\abs{s}$ denotes Hamming weight. Note in particular that the empty partition is represented by $\ket{\emptyset}=\ket{1^k0^{n-k}}$.  Now,  let$ X_{\lambda,\emptyset}$ denote the $k\times k$ submatrix obtained by selecting the first $k$ columns, selected by $\ket{\emptyset}$ and the $k$ rows determined by $\lambda$. Note that all these  depend only on the subset of variables $(x_1,\ldots, x_k)$. Then, the  claim is that
\equ{
s_{\lambda}(x_1,x_2,\ldots,x_k) = \frac{1}{V_k(x)}\det X_{\lambda,\emptyset}\,.
}
where $V_k(x)=\prod_{1\leq i<j\leq k}(x_j-x_i)$.  Note that selecting any other set of $k$ columns amounts to a simple relabeling of the variables and thus there is no loss of generality in taking the first $k$ columns in this definition.\footnote{The Schur functions $s_\lambda (x_1,\cdots, x_{n})$ have a “stability” property. For any $k\leq n$, i.e., $s_\lambda (x_1,\cdots, x_{k},0,\ldots0)=s_\lambda (x_1,\cdots, x_{k})$.} Also note that for $k=n$, $P_{n,n}=\{\emptyset\}$ consists only of the empty partition, in which case one obtains $s_\emptyset(x_1,x_2,\ldots,x_n)=V_n^{-1}\det X^{(n)}=1$, consistent with the general definition.

\section{Computational complexity}
\label{app:Computational complexity}

Here we summarize some relevant results on the computational complexity of various coefficients. See \cite{panova2023computational,panova2023complexity} for a  recent review and open problems.  A unifying formalism for discussing various coefficients is that of plethysm.  The Littlewood-Richardson, Kronecker, and plethysm coefficients arise from the following expansions \cite{stanley1999enumerative,macdonald1998symmetric,stanley2000positivity}: 
\eqs{
s_\mu[X+ Y]=\,& \sum_{\lambda, \nu} C_{\nu\lambda}^{\;\;\;\;\mu}s_{\nu}[X]s_{\lambda}[Y]\,,\\
s_\mu[X Y]=\,& \sum_{\lambda, \nu} g_{\mu\nu \lambda}s_{\nu}[X]s_{\lambda}[Y]\,,\\
s_\mu[s_{\nu}[X]]=\,& \sum_{\lambda} p_{\mu\nu \lambda}s_{\nu}[X]s_{\lambda}[X]\,,
}
where $[X+Y]$ and $[XY]$ represent the plethysm union and composition.  Recall that Schur functions are characters of the general linear group, $s_\lambda=\Tr \rho_\lambda$, with the $x$’s corresponding to the eigenvalues of $\rho_{\lambda}$. Then, the first and third formulas above follow from those in representation theory. There are a number of open problems related to all the coefficients above; see Stanley’s list of problems in \cite{stanley2000positivity}. The Littlewood-Richardson coefficients are well understood combinatorially. The Kronecker coefficients less so and the plethysm coefficients are the most elusive of all (see, e.g., \cite{colmenarejo2022mystery}).

\begin{table}[]
\begin{center}
\begin{tabular}{|c|c|c|c|} \hline
&Decision $>0$ & Exact $\#$ & Approx $\#$   \\ \hline
LR (binary)& $\P$ \cite{burgisser2013deciding} & $\#\P$-complete \cite{narayanan2006complexity}   & $?<\mathsf{LR}\leq \BPP^{\NP}$     \\ \hline
Kronecker (unary) & \cite{Ikenmeyer_2017} $\NP\leq \mathsf{Kron}\leq \QMA$ \cite{bravyi2023quantum} & \cite{burgisser2008complexity}  $\#\P\leq \mathsf{Kron}\leq \#\BQP$ \cite{bravyi2023quantum}    & $\BPP^\NP<\mathsf{Kron}\leq \QXC$  \cite{bravyi2023quantum}   \\ \hline
Plethysm  & \cite{fischer2020computational} $\NP\leq \mathsf{Pleth}<?$ &  \cite{fischer2020computational}  $\#\P\leq \mathsf{Pleth}\leq \GapP$   \cite{fischer2020computational}   & $\BPP^\NP\leq \mathsf{Pleth}\leq \#\P$    \\  \hline \hline
Kostka (binary) & $\P$ \cite{panova2023computational} & $\#\P$-complete  \cite{narayanan2006complexity}   & $?<\mathsf{Kostka}\leq \BPP^{\NP}$  \\ \hline
\end{tabular}
\caption{The complexity of deciding whether the coefficient is positive, its exact computation, and a multiplicative approximation. Recall that $\mathsf{C_=P}=\coNQP$. Note the containment results for Kronecker in unary do {\it not} imply containment in binary but the hardness results for unary do imply hardness for binary.  }
\end{center}
\label{table:LKPcomp}
\end{table}%

\paragraph{Note:} The inputs can be provided either in binary or unary. Hardness in unary implies hardness in binary. Containment in binary implies containment in unary. The converse of these statements does not hold. 
\paragraph{Littlewood-Richardson coefficients.} The  problem of deciding whether $C_{\mu\nu}^{\;\;\;\;\lambda}>0$ is in $\P$  \cite{burgisser2013deciding}, based on the formulation of Littlewood-Richardson coefficients  by Knutson and Tao as counting the number of solutions to certain honeycomb puzzles. The exact computation of Littlewood-Richardson coefficients was shown to be $\#\P$-complete when the input is in binary by Narayanan \cite{narayanan2006complexity}. See \cite{panova2023computational} for the case of unary input.

\paragraph{Kostka numbers.} These are similar to the Littlewood-Richardson coefficients: the problem of deciding whether the Kostka numbers are $K_{\mu\nu}>0$ is in $\P$  \cite{panova2023computational} for unary (and hence binary) inputs.  The counting is $\#\P$-complete \cite{narayanan2006complexity}.  for binary inputs and it is conjectured to be so for unary inputs a well (see conjecture 5.11 in \cite{panova2023computational}).

\paragraph{Kronecker Coefficients.} Deciding positivity of the Kronecker coefficients is $\NP$-hard \cite{Ikenmeyer_2017}. It is not known whether the problem is in $\NP$, but it has been shown to be in $\QMA$  \cite{bravyi2023quantum}. On the exact computation, it is suspected that the problem is {\it not} in $\#\P$ (see open Problem 5.16 in \cite{panova2023computational}). When the input is in unary, the results of \cite{bravyi2023quantum} show that it is contained in $\#\BQP$.  It is not known whether the containment results in  \cite{bravyi2023quantum} (in $\QMA$ for the decision problem and  in $\#\BQP$ for the counting problem) hold for  {\it binary} inputs (see open Problem 5.21 in  \cite{panova2023computational}).

\paragraph{Plethysm Coefficients.}  The problem of deciding whether $p_{\mu\nu\lambda}>0$ is $\NP$-hard  and computing them exactly is $\#\P$-hard  \cite{fischer2020computational} and contained in $\GapP$. It is not known whether they are in $\#\P$ and it is suspected they are not  \cite{panova2023computational}.

\paragraph{Characters of $S_n$.} The problem of deciding if $\chi_{S_n}=0$ is $\C_=\P$-complete \cite{ikenmeyer2022positivity}. It is also shown that computing the (square) of the characters of $S_n$, given $\mu, \lambda\vdash n$ as input, {\it cannot} be in $\#\P$, unless the polynomial hierarchy collapses to the second level. It also follows from the proof that deciding positivity of the character (not its square) is PP-complete under Karp reductions, and hence $\PH$-hard under  Turing reductions. It is conjectured there that under Karp reductions the decision problem is $\PP$-complete and the counting version is $\GapP$-complete. See Table~\ref{tab:charactercom} for a summary. 
 
\begin{table}[]
\begin{center}
\begin{tabular}{|c|c|c|c|} \hline
&Decision $>0$ & Exact $\#$ & Approx $\#$   \\ \hline
LR (unary) & $\P$ \cite{panova2023computational} & $\#\P$-complete? (conjectural \cite{panova2023computational})   & $?<\mathsf{LR}\leq \BPP^{\NP}$     \\ \hline
Kostka (unary) & $\P$ \cite{panova2023computational} & $\#\P$-complete? (conjectural \cite{panova2023computational})   & $?<\mathsf{Kostka}\leq \BPP^{\NP}$  \\ \hline
\end{tabular}
\end{center}
\caption{Conjectures on the hardness in unary representation. Note these would imply the known hardness results in binary. }
\label{default}
\end{table}%

\begin{table}[]
\begin{center}
\begin{tabular}{|c|c|c|c|} \hline
&Decision & Exact $\#$ & Approx $\#$   \\ \hline
$\chi^2_{S_n}$ (Karp) & $\mathsf{C_=P}$-complete \cite{ikenmeyer2022positivity} &  \cite{ikenmeyer2022positivity} $\#\P<\chi^2_{S_n}\leq \GapP$  \cite{ikenmeyer2022positivity} &  $\BPP^\NP<\chi^2_{S_n}\leq $\#\P$$   \\ \hline 
$\chi_{S_n}$  (Turing)  & $\PH$-hard  \cite{ikenmeyer2022positivity} &   $\GapP$-complete \cite{ikenmeyer2022positivity}  &  $\#\P$-complete   \\ \hline \hline
$\chi_{S_n}$ (Karp) & $\PP$-complete  \cite{ikenmeyer2022positivity} &  $\GapP$-complete? (conjectural) \cite{ikenmeyer2022positivity}  &  $\#\P$-complete? (conjectural)   \\ \hline
\end{tabular}
\caption{The complexity of decision and exact computation of the character and character squared of $S_n$, under  polynomial time (i.e. Karp) and Turing (i.e. Cook or oracle) reductions. Hardness under Karp reductions implies hardness under Turning reductions, but not the other way around.}
\label{tab:charactercom}
\end{center}
\end{table}%

\end{document}